\documentclass[11pt]{article}
\usepackage{amsmath,amsbsy,amsfonts,amssymb,graphics,epsfig,float,mathrsfs,amsthm,braket}
\usepackage{ifsym}
\usepackage{psfrag}
\usepackage{color}
\usepackage[normalem]{ulem}
\usepackage{multirow}

\newcommand{\error}{\mathrm{err}}
\newcommand{\beq}{\begin{equation}}
\newcommand{\eeq}{\end{equation}}
\newcommand{\beqst}{\begin{equation*}}
\newcommand{\eeqst}{\end{equation*}}
\newcommand{\tp}{\tilde{P}}
\newcommand{\dmin}{\delta_{\mathrm{min}}}
\newcommand{\dmax}{\delta_{\mathrm{max}}}
\newcommand{\tdmax}{\tilde{\delta}_{\mathrm{max}}}
\newcommand{\tdmin}{\tilde{\delta}_{\mathrm{min}}}
\newcommand{\ddelta}{\Delta \delta}
\newcommand{\dtdelta}{\tilde{\Delta \delta}}

\newcommand{\ho}{h_0}
\newcommand{\Tpre}{T_{\mathrm{pre}}}
\newcommand{\TpreEff}{T_{\mathrm{eff}}}

\newcommand{\Rf}{R_{\mathrm{clamp}}}
\newcommand{\Rcurv}{R_{\mathrm{curv}}}
\newcommand{\Ro}{\Rf}
\newcommand{\Ri}{R_{\mathrm{in}}}
\newcommand{\rin}{\rho_{\mathrm{in}}}
\newcommand{\eqq}{\epsilon_{\theta\theta}}

\newcommand{\sqq}{\sigma_{\theta\theta}}
\newcommand{\srr}{\sigma_{rr}}

\newcommand{\upd}{\mathrm{d}}
\newcommand{\tdelta}{\tilde{\delta}}
\newcommand{\tPhi}{\tilde{\Phi}}
\newcommand{\cF}{{\cal F}}
\newcommand{\Fsum}{{\cal F}_{\mathrm{sum}}}
\newcommand{\Fbend}{F_{\mathrm{bend}}}
\newcommand{\Fmem}{F_{\mathrm{mem}}}
\newcommand{\lec}{\ell_{ec}}

\newcommand{\s}[1]{{\textsf{\textbf{#1}}}}

\begin{document}
\title{\s{Indentation metrology of clamped, ultra-thin elastic sheets}}
\author{ \textsf{Dominic Vella\textit{$^{a}$} and Benny Davidovitch\textit{$^b$}}\\ 
{\it$^a$Mathematical Institute, University of Oxford, Woodstock Rd, OX2 6GG, UK}\\
{\it$^{b}$Department of Physics, University of Massachusetts Amherst, Amherst, MA 01003, USA}}

\date{\today}
\maketitle
\hrule\vskip 6pt
\begin{abstract}
We study the indentation of ultrathin elastic sheets clamped to the edge of a circular hole. This classical setup has received considerable attention lately, being used by various experimental groups as a probe to measure the surface properties and stretching modulus of thin solid films. Despite the apparent simplicity of this method, the geometric nonlinearity inherent in the mechanical response of thin solid objects renders the analysis of the resulting data a  nontrivial task. Importantly, the essence of this difficulty is in the geometric coupling between in-plane stress and out-of-plane deformations, and hence is present in the behaviour of Hookean solids even when the slope of the deformed membrane remains small. Here we take a systematic approach to address this problem, using the membrane limit of  the F\"{o}ppl-von-K\'{a}rm\'{a}n equations. This approach highlights some of the dangers in the use of approximate formulae in the metrology of solid films, which can introduce large errors; we suggest how such errors may be avoided in performing experiments and analyzing the resulting data.
\end{abstract}
\vskip 6pt
\hrule

\maketitle

\section{Introduction}

Indentation experiments have  recently become a popular tool with which to characterize the mechanics of thin solid sheets at a variety of length scales: from the stretching modulus of the thinnest known material --- graphene \cite{Lee2008} --- to the ``pre-tension" and  surface energy of polymer films with thicknesses $t$ in the range  $10\mathrm{~nm}\lesssim t\lesssim 10^3\mathrm{~nm}$\cite{Xu2016,Wan2003}. With the advance of technology, and the ability to induce highly controlled and localized forces with micro- and nano-transducers, as well as AFM devices, indentation experiments can be preformed with a high level of precision, providing detailed, highly valuable data for metrology. 
However, the subtle nature of the elasticity of solid sheets, which intertwines geometry and mechanics in a complex manner, renders the analysis and interpretation of the resulting experimental data a nontrivial task.  These subtleties have led to a good deal of confusion, including the inappropriate application of asymptotic results as well as the propagation of over-simplified \emph{ad-hoc} formulae. In this paper we aim to give a clear description of the underlying mechanics, together with rationalizations of a number of asymptotic results, and some simple demonstrations of the pitfalls of misusing these results.

\subsection{Background\label{sec:background}}

\begin{figure}
\centering
\includegraphics[width=0.6\columnwidth]{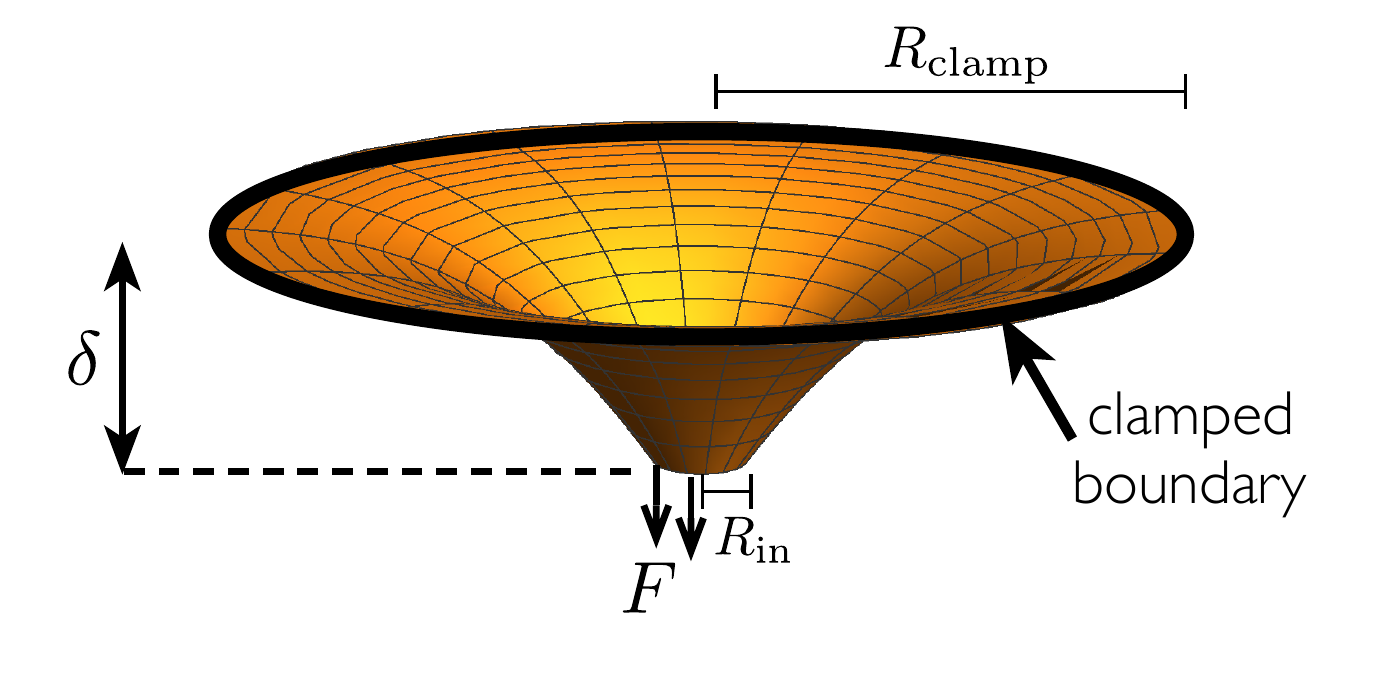}
\caption{Schematic sketch of the indentation of a clamped membrane. An indentation, depth $\delta$ is imposed at the centre and requires the application of a force $F$. }
\label{fig:setup}
\end{figure}

A typical indentation setup (fig.~\ref{fig:setup}) consists of a  circular sheet, strongly attached to a planar rigid rim of radius $\Rf$. An indenter with a sharp tip (modelled here as a disk of radius $\Ri \ll \Rf$) then exerts a normal load within the region $r\leq \Ri$. The edge $r=\Rf$ is assumed to be effectively {\emph{clamped}} to the rim, preventing it from sliding inward in response to indentation.  For very thin sheets, the effect of  bending stiffness may be neglected. This already idealized problem may be simplified further by approximating the indenter by a point force, \emph{i.e.}~$\Ri=0$, and assuming that the sheet is not subject to any tension in its clamped, undeformed state. This highly simplified problem is the classic problem that was considered by Schwerin \cite{Schwerin1929}, who found that the force $F$ is cubic in the imposed deflection, $\delta$ (see fig.~\ref{fig:setup}). Schwerin's result may be rationalized as the balance between the work done by the indenter, $F\delta$, and the stretching energy in the sheet, $Y\Rf^2\epsilon^2$ where the typical strain $\epsilon\sim (\delta/\Rf)^2$ and $Y=Et$ is the stretching modulus of the sheet, with $E$ and $t$ the Young's modulus and thickness of the sheet, respectively.

The scaling version of Schwerin's result may be generalized to the case in which the sheet is subject to a uniform tension $\Tpre$ in its clamped state. The energy balance becomes
\beq
F\cdot \delta \sim\Rf^2 \left[\Tpre + Y \left(\frac{\delta}{\Rf}\right)^2 \right]  \left(\frac{\delta}{\Rf}\right)^2  \ , 
\label{Eq:scaling1}
\eeq  
since the relevant stress within the sheet  is now estimated as the sum of the pre-tension $\Tpre$ and the typical poking-induced stress, $Y (\delta/\Rf)^2$. For $\Tpre >0$, Eq.~(\ref{Eq:scaling1}) suggests that upon increasing $\delta$, the force transforms from a linear response, $F/\delta \sim \Tpre$, to the nonlinear Schwerin response, $F / \delta^3  \sim Y/\Rf^2$, depending on the indentation depth $\delta$. Indeed, noting that the transition between these qualitatively distinct responses occurs at $\delta \sim \Rf (Y/\Tpre)^{1/2}$, it is useful to define the dimensionless  indentation depth: 
\beq
\tdelta=\frac{\delta}{\Rf}\left(\frac{Y}{\Tpre}\right)^{1/2}.
\label{eqn:DispND}
\eeq 
The transition from linear to cubic force--displacement laws is therefore expected to occur when $\tdelta \sim O(1)$. It is important to notice that, if the stretching modulus is sufficiently large in comparison to the pre-tension (as in most indentation experiments), it is easy to reach the nonlinear regime ($\tdelta \gg1$) even while the characteristic slope ($\sim \delta/\Rf$) remains very small. This observation means that it is appropriate to use the F\"{o}ppl-von-K\'{a}rm\'{a}n (FvK) equations\cite{Landau,Mansfield}, which assume small slope, but nevertheless do capture the geometric nonlinearity of the response.    

It is also useful to introduce a dimensionless measure of the force: 
\beq
\cF=\frac{Y^{1/2}F}{\Tpre^{3/2} \Rf} \ .
\label{eqn:ForceND}
\eeq With this definition, Schwerin's law may be recast in the simple form: $\cF/\tdelta^3 = \alpha(\nu)$, where $\nu$ is the Poisson ratio, and $\alpha(\nu)$ is a smooth, nearly constant function, which has  been computed previously \cite{Begley2004}. 

\subsection{Main results}
At a qualitative level of understanding, Eq.~(\ref{Eq:scaling1}) describes correctly the nature of the force-displacement function. From an experimental point of view, however, what is desired is an analytical formula for $\cF(\tdelta$) that interpolates correctly between the linear and nonlinear regimes, and is uniformly valid over the whole interval of feasible indentation depths. Unfortunately, such a formula does not exist, even while restricting attention to purely Hookean responses. As we will show in this paper, this difficulty is intimately related to the ideal nature of Schwerin's calculation, which ignores both pre-tension and the finite size of the indenter. We find that Schwerin's ideal approach is a useful starting point in the nonlinear regime, $\tdelta \gg 1$, where the effects of $\Tpre$ and $\Ri$ can be accounted for as regular perturbations of Schwerin's result. However, for small indentation depths, $\tdelta \ll 1$, the effects of both pre-tension and the finite size of the indenter tip are singular and intertwined. In particular,  we find a linear response with a spring constant, $\cF/\tdelta \approx 2\pi/\log(\Rf/\Ri)$, that vanishes as $\Ri \to 0$. However, for a point indenter, the linear response with $\tdelta \ll 1$  becomes sub-linear so that $\cF/\tdelta\to0$ as $\tdelta\to0$.  Our results on the effect of the size of the indenter tip are summarized in Table I.  

One notable example, to which our results should be particularly relevant, is the celebrated measurement of the stretching modulus of graphene \cite{Lee2008}; here an experimental force--indentation curve, $F(\delta)$, was obtained by using an Atomic Force Microscope (AFM) as the indenter. Following \cite{Begley2004},  Lee \emph{et al.}\cite{Lee2008} assumed that $F(\delta)$ can be expressed as an algebraic sum of Schwerin's nonlinear term and a linear term whose coefficient is proportional to some unknown tension $\Tpre$, independently of the indenter's size. Fitting this proposed algebraic expression (with 2 unknown parameters) to the measured $F(\delta)$, the authors evaluated the pre-tension, and the stretching modulus $Y$. In \S\ref{sec:finite} we discuss the accuracy of this approach, and  show that it may often lead to significant errors in the estimated values of the stretching modulus and pre-tension in the sheet.  Furthermore, in \S\ref{sec:pres} we propose a method to extract the stretching modulus from the linear regime of small indentation depth for sheets subject to a large pressure (a `nano-balloon').    

 
In the polymer science community, several workers have used various approaches to describe the metrology of thin polymer sheets from indentation measurements. However, these works often use uncontrolled (and/or over-simplified) assumptions, or include unnecessary details in the model:  

{\emph{(A)}} One example is ref.~\cite{Wan2003}, in which the stress in the sheet is assumed to be uniform and isotropic throughout the indentation (though increasing with indentation depth). This simplification facilitates analytical progress but neglects an important difference between solid sheets and liquid membranes (which cannot support anisotropic stresses in equilibrium). Though the scaling behaviour that results from such analyses is correct, the calculated prefactors can vary considerably \cite{Mitchell2003}, undermining the validity of any resulting fit. 



{\emph{(B)}} Several previous works have provided numerically-determined plots of the force--displacement relationship, together with the appropriate asymptotic limits of this force--displacement relationship in the limits of large and small indentation depths, as discussed in \S\ref{sec:background}. While these calculations are  correct, the authors of these studies often present approximate analytical formulae obtained by adding the two asymptotic results (an additive composite expansion \cite{Hinch1990}). However, they report these analytical formulae without any discussion of the errors inherent in their use. We shall show that the errors introduced can be large, particularly at the  intermediate indentation depths that are often encountered experimentally.

{\emph{(C)}} An unnecessary complication in numerous models of indentation is the inclusion of bending forces \cite{Wan2003,Xu2016}. As we will show, despite enhancement of bending forces by strong spatial variation of the profile, they are dwarfed by tensile forces, and  can be safely ignored in many experimentally-relevant situations. 

{\emph{(D)}} Another flawed approach for indentation-assisted metrology,  proposed very recently \cite{Xu2016}, is to extract the pre-tension from the deformed shape of the sheet. As we  show below, the only robust information that can be extracted by fitting the shape is whether the response is dominated by pre-tension ({\emph{i.e.}} $\tdelta \ll 1$), or rather by the stretching of the sheet ({\emph{i.e.}} $\tdelta \gg 1$). However, any attempt to determine the actual value of the pre-tension from measurements of the shape is doomed to fail.  

\subsection{Outline}
We start in \S\ref{sec:goveq} by setting up the equations, identifying dimensionless groups that govern the mechanics, and performing a simple calculation that reveals the singular nature of the linear response regime. In \S\ref{sec:point} we specialize to the case of pointwise indentation, for which we obtain an analytical solution, valid for the whole range of indentation amplitude, and show that the regime $\tdelta \ll 1$ is characterized by a sub-linear response. In \S\ref{sec:finite} we return to an indenter of finite size, $\Ri \ll \Rf$, and characterize the singular nature of the linear response at $\tdelta \ll 1$, together with its relationship to the point-indenter results. We then discuss, in \S\ref{sec:pres}, how an internal pressure affects these results. In \S\ref{sec:crit} we use our results to critique previous works and shed light on some subtleties and sources of confusion in this problem. Finally, in \S\ref{sec:conclude} we conclude and note an important effect on the response if the clamped boundary conditions are relaxed.

\begin{table*}[h]
\small
  \caption{\ A summary of the main results for the force--displacement relations}
  \label{tbl:example}
  \begin{tabular*}{\textwidth}{@{\extracolsep{\fill}}llll}
    \hline
   \multirow{2}{*}{Indenter Size} & Small & Intermediate & Large \\
   & displacements & displacements & displacements\\
    \hline
    $\Ri=0$ & $\cF/\tdelta \approx 2\pi/\log(4/\tdelta)$ & ---  & $\cF/\tdelta^3 \approx\alpha(\nu)$  (Schwerin\cite{Schwerin1929})\\
    \hline
    $\Ri\ll\Rf$ & $\cF/\tdelta \approx 2\pi/\log(\tfrac{\Rf}{\Ri})$  & $\cF/\tdelta \approx 2\pi/\log(4/\tdelta)$  & 
    $\cF/\tdelta^3 \approx\alpha(\nu) + O(\Ri/\Rf)^{2/3} $ \\
    \hline
  \end{tabular*}
\end{table*}

\section{The FvK equations\label{sec:goveq}}

We begin with the F\"{o}ppl-von-K\'{a}rm\'{a}n (FvK) equations \cite{Landau,Mansfield} relating the out-of-plane membrane displacement $\zeta(r)$ to the Airy stress function $\psi(r)$. (Here $\psi$ is defined such that the principal stresses are $\srr=\psi/r$ and $\sqq=\psi'$, where we use the axial symmetry of the setup.) We then have the vertical force balance equation for the membrane in $\Ri<r<\Rf$, \emph{i.e.}
\beq
-\frac{1}{r}\frac{\upd}{\upd r}\left(\psi\frac{\upd \zeta}{\upd r}\right)=0
\label{eqn:FvK1}
\eeq and hence
\beq
\psi\frac{\upd\zeta}{\upd r}=\frac{F}{2\pi}
\label{eqn:vfbal}
\eeq where the constant of integration is related to the indentation force $F$ applied via a simple force balance.

The in-plane stress is coupled to the out-of-plane displacement by the compatibility of strains equation, i.e.
\beq
r\frac{\upd}{\upd r}\left[\frac{1}{r}\frac{\upd}{\upd r}\left(r\psi\right)\right]=-\tfrac{1}{2}Y\left(\frac{\upd\zeta}{\upd r}\right)^2.
\label{eqn:compat}
\eeq 

\subsection{Boundary conditions}

The governing equations \eqref{eqn:vfbal}--\eqref{eqn:compat} are to be solved with appropriate boundary conditions. The conditions on the vertical displacement are clearly:
\beq
\zeta(\Ri)=-\delta,\quad\zeta(\Rf)=0,
\label{eqn:indBC}
\eeq corresponding to an imposed indentation depth at the indenter and zero vertical displacement (where the clamping is imposed) at the outer edge of the film, $r=\Rf$.

The clamping boundary condition requires a little thought. Prior to clamping and additional deformation being imposed, there is a base horizontal displacement (due to the pre-tension $\Tpre$) $u(r)=u_0(r)$, where
\beq
\frac{u_0(r)}{r}=\eqq=\frac{\sqq-\nu\srr}{Y}=\frac{(1-\nu)\Tpre}{Y},
\label{eqn:clamping}
\eeq with $Y=Et$ the stretching modulus of the material and $\nu$ its Poisson ratio. We assume that this clamping is imposed, and is perfectly effective, at both the outer edge of the film and the point where contact is first made with the indenter (corresponding to a cylindrical, no-slip indenter). Other variants of this condition, \emph{e.g.}~perfect slip, are expected only to modify the numerical pre-factors in the analysis that follows.

At the points where clamping in imposed, the horizontal displacement must remain at the original values given by $u_0(r)$, i.e.
\beqst
\frac{u(\Rf)}{\Rf}=\frac{u(\Ri)}{\Ri}=\frac{(1-\nu)\Tpre}{Y}.
\eeqst Since our problem is most commonly solved in terms of the stress within the film using the Airy stress function $\psi(r)$, it is useful to express the clamping boundary condition as: 
\beq
\psi'(\Ri)-\nu\frac{\psi(\Ri)}{\Ri}=\psi'(\Rf)-\nu\frac{\psi(\Rf)}{\Rf}=(1-\nu)\Tpre.
\label{eqn:stressBC}
\eeq

\subsection{Non-dimensionalization}

To facilitate the solution of the problem, we use  dimensionless variables in the remainder of the paper, letting
\beq
\rho=r/\Rf,\quad \Psi=\psi/(\Tpre\Rf),\quad Z=\frac{\zeta}{\Rf}\left(\frac{Y}{\Tpre}\right)^{1/2}. 
\label{eqn:NonDim}
\eeq The dimensionless versions of the governing equations \eqref{eqn:vfbal}--\eqref{eqn:compat} and boundary conditions \eqref{eqn:indBC}--\eqref{eqn:stressBC} are given in Appendix A.

Our problem depends on three dimensionless parameters. The first is the geometric parameter $$\rin=\Ri/\Rf,$$ which measures the radius of the indenter to that of the membrane. To simplify the discussion we will assume that $\rin$ is constant throughout a particular experiment (i.e.~the indenter is cylindrical). However, we note in passing that, for a non-cylindrical indenter,  $\rin$, may depend on $\delta$; we shall discuss the significance of this in light of our results in \S\ref{sec:crit:small}.

The other two dimensionless parameters, defined in \eqref{eqn:DispND} and \eqref{eqn:ForceND}, evolve during the indentation: the dimensionless indentation depth $\tdelta$, \eqref{eqn:DispND}, gives a measure of the indentation depth compared to that at which the stress induced by indentation becomes comparable to the pre-tension. We therefore expect that for $\tdelta\ll1$ the tension in the membrane is `close' to the pre-tension (with caveats that we discuss in due course); for $\tdelta\gg1$ the effect of the pre-tension is expected to be negligible. The final dimensionless parameter is the dimensionless indentation force $\cF$,  \eqref{eqn:ForceND}. The key quantity of interest is therefore the dimensionless force--displacement relationship $\cF(\tdelta)$.

In the limit of small indentations, $\tdelta\ll1$, the fact that the tension is approximately unchanged from the state prior to indentation may be exploited to show that the poking force $\cF$ is linear in $\tdelta$ (see Appendix A and Jennings \emph{et al.}\cite{Jennings1995}); in dimensionless terms:
\beq
\cF=\frac{2\pi}{\log(1/\rin)}\tdelta.
\label{eqn:FLawRin}
\eeq This simple force law shows that the limit $\rin\to0$ is singular: apparently the membrane becomes arbitrarily compliant for sufficiently small $\rin$ and $\tdelta$. To understand better what happens as $\rin\to0$ we first consider the limit $\rin=0$, a point indenter. We shall  see that for $\rin\ll1$ the small indentation behaviour  \eqref{eqn:FLawRin} only holds for $\tdelta\lesssim\tdelta_\ast(\rin)$, where  $\tdelta_\ast(\rin)\to0$ as $\rin\to0$. Instead, a new response emerges for $\tdelta_\ast(\rin)\lesssim\tdelta\ll1$ that is independent of the indenter radius.

\section{Point Indentation\label{sec:point}}

With $\rin=0$, the problem simplifies considerably, allowing analytical progress to be made; the details of this analytical calculation are presented in Appendix B.

\subsection{Analytical results\label{sec:point:analysis}}

We are able to find (see Appendix B) a parametric representation for the force--displacement relationship in terms of $\tPhi_1(\tdelta)$, which is defined implicitly by
\beq
\tdelta(\tPhi_1)=\frac{2}{A(\tPhi_1)^{1/2}}\sinh^{-1}(\tPhi_1^{1/2}) 
\label{eqn:tdeltaPar}
\eeq where
\beq
A(\tPhi_1)=\frac{2}{1-\nu}\left[1-\left(\frac{1+\tPhi_1}{\tPhi_1}\right)^{1/2}\sinh^{-1}(\tPhi_1^{1/2})\right]+\tPhi_1.
\label{eqn:Apar}
\eeq The indentation force required to obtain a particular indentation depth is then given by
\beq
\cF(\tPhi_1) =\frac{4\pi}{A(\tPhi_1)^{3/2}}\left[\tPhi_1^{1/2}(1+\tPhi_1)^{1/2}-\sinh^{-1}(\tPhi_1^{1/2})\right]
\label{eqn:cFPar}
\eeq with $\tPhi_1$ and $A(\tPhi_1)$ as defined in \eqref{eqn:tdeltaPar} and \eqref{eqn:Apar}, respectively.

\begin{figure}
\centering
\includegraphics[width=0.6\columnwidth]{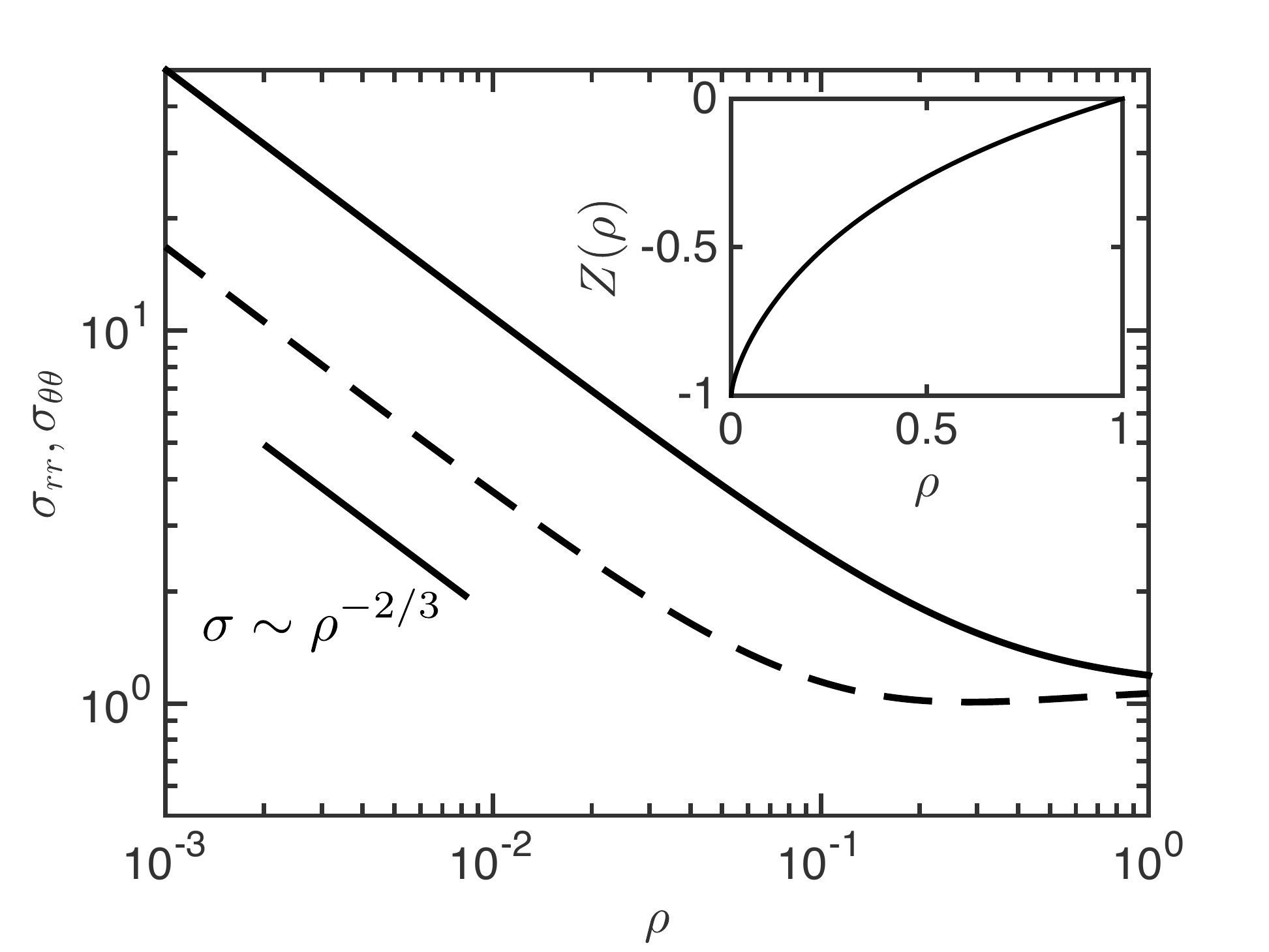}
\caption{Profiles from the analytical solution for a point indenter with indentation depth $\tdelta=1$, $\nu=1/3$. Main figure: the principal stresses $\srr$ (solid curve) and $\sqq$ (dashed curve), show a $\rho^{-2/3}$ (solid line) singularity as the indenter is approached, $\rho\to0$. Inset: The profile of the indented sheet.}
\label{fig:PointProfiles}
\end{figure}

Analytical expressions may also be found for the stress within the membrane, and the vertical displacement of the membrane (see Appendix B). Typical profiles for the stress and  membrane displacement are shown in figure \ref{fig:PointProfiles}.

%

\subsection{Force law}

In Eqs \eqref{eqn:tdeltaPar} and \eqref{eqn:cFPar}, we have a complete parametric representation of the force--displacement relationship. It is, however, useful to extract from this exact relationship simple, approximate force laws that may be used in the limits of small and large indentation depths.

\begin{figure}
\centering
\includegraphics[width=0.6\columnwidth]{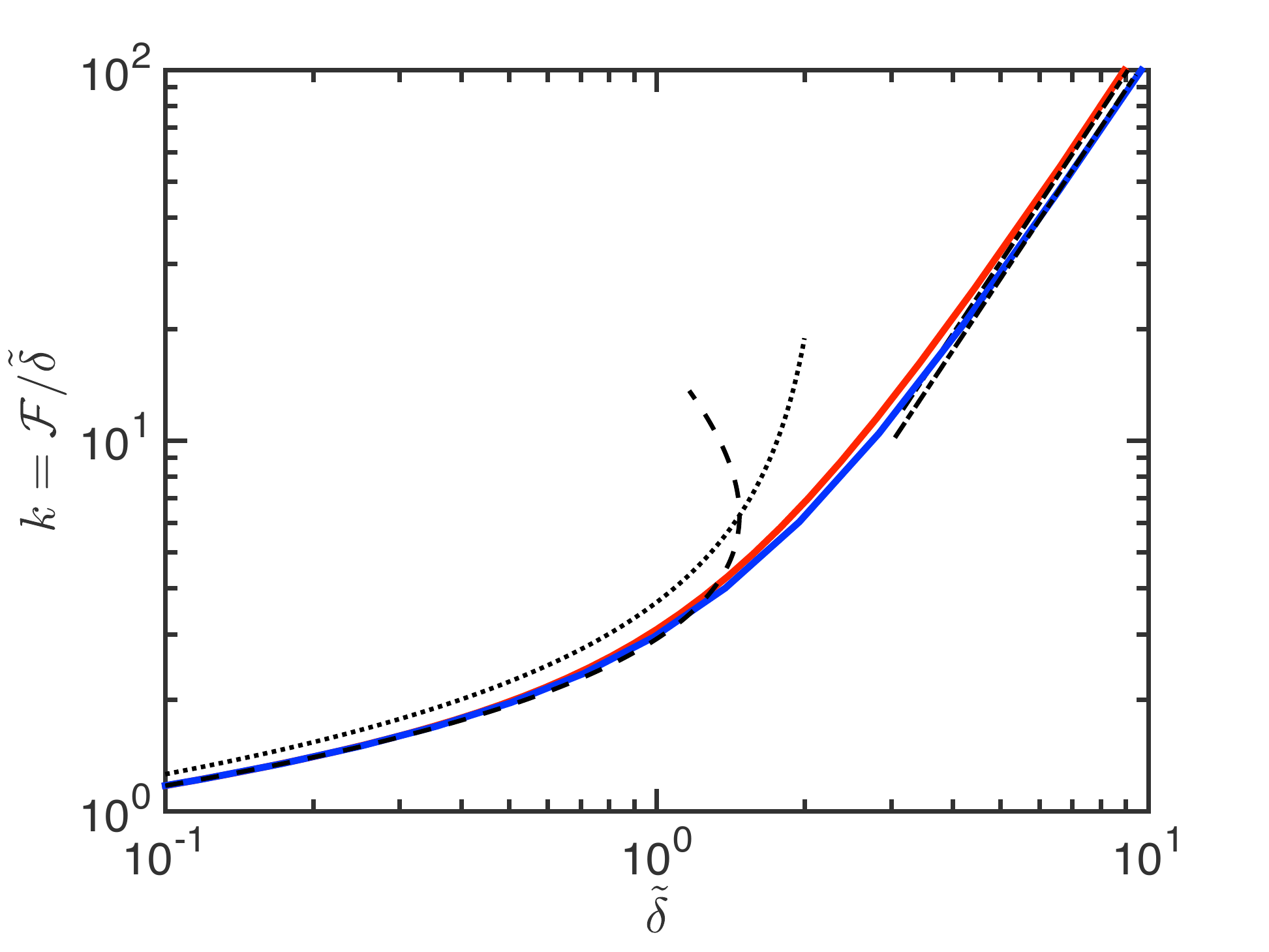}
\caption{The numerically-determined stiffness, $k=\cF/\tdelta$, for point indentation of a clamped membrane. Solid curves show the full result (obtained by plotting $\tdelta(\tPhi_1)$ and $\cF(\tPhi_1)$ parametrically from  \eqref{eqn:tdeltaPar}--\eqref{eqn:cFPar}, respectively. Results are shown for $\nu=1/2$ (red) and $\nu=1/3$ (blue). The plot of stiffness versus $\tdelta$ emphasizes that, in the point indenter limit, there is no true linear stiffness (\emph{i.e.}~there is no region in which $\cF\propto\tdelta$). Asymptotic results are also shown: \eqref{eqn:ForceLawSmallD} (dashed curve) is valid for $\tdelta\ll1$ while \eqref{eqn:ForceLawLargeD} (dash-dotted lines) is valid for $\tdelta\gg1$. Note also that the explicit form of the small displacement force law \eqref{eqn:ForceLawInvert} (dotted curve), gives less satisfactory agreement with the analytical result than the implicit form \eqref{eqn:ForceLawSmallD}.}
\label{fig:ForceLawPoint}
\end{figure}

For small indentation depths, $\tdelta\ll1$, we find\footnote{A similar result was obtained in a simplified model with a constant applied tension \cite{Norouzi2006}.} that the dimensionless indentation force $\cF$ satisfies
\beq
\tdelta=\frac{\cF}{2\pi}\log(8\pi/\cF),
\label{eqn:ForceLawSmallD}
\eeq which can be approximately inverted to give
\beq
\cF=\frac{2\pi\tdelta}{\log\left[\frac{4}{\tdelta}\log\frac{4}{\tdelta}\right]}.
\label{eqn:ForceLawInvert}
\eeq (However, note that the original form in \eqref{eqn:ForceLawSmallD} has an appreciably smaller error when compared with the full result, as shown in figure \ref{fig:ForceLawPoint}.) The force law in \eqref{eqn:ForceLawInvert} shows sub-linear growth with indentation depth, $\tdelta$: $\cF/\tdelta\to0$ as $\tdelta\to0$. For a finite indenter radius, $\rin$, \eqref{eqn:ForceLawInvert} therefore represents a softer spring than the linear force law given in \eqref{eqn:FLawRin}. However, in the limit $\rin\to0$, the point-like response \eqref{eqn:ForceLawInvert} prevents the arbitrary softening that led us to consider the point indenter limit. 

For large indentations, we find that
\beq
\cF=\alpha(\nu)\tdelta^3
\label{eqn:ForceLawLargeD}
\eeq where $\alpha(\nu)$ is a prefactor that must be determined from the solution of a transcendental equation, see Appendix B. Crucially, we find  that $\alpha(\nu)$ does not vary significantly with Poisson ratio $\nu$ in the relevant range: $\alpha(1/3)=\pi/3\approx1.047$ and $\alpha(1/2)\approx1.213$. The approximation $\alpha(\nu)\approx0.867+0.2773\nu+0.8052\nu^2$ is accurate to within $0.7\%$ for all $0\leq\nu\leq1/2$.

The force--displacement relationship given by \eqref{eqn:tdeltaPar} and \eqref{eqn:cFPar} is shown in figure \ref{fig:ForceLawPoint} for two different values of the Poisson ratio $\nu$, together with the asymptotic results given above. We see that the agreement is good and, further, that, as expected from the hypothesized independence of $\Tpre$, $\cF\sim\tdelta^3$ for $\tdelta\gg1$. However, the key observation is that at small indentation depths the force law is subtly different from the linear relation $F\propto\delta$ that is often assumed \cite{Schwerin1929,Begley2004,Komaragiri2005,LopezPolin2015}. The result, \eqref{eqn:FLawRin},  corresponds to an apparent stiffness that increases logarithmically with increasing indentation (as seen in figure \ref{fig:ForceLawPoint}). Its appearance is intimately related to the point indenter assumption, since this causes both components of the stress to grow indefinitely as the origin is approached, $\srr,\sqq\propto\rho^{-2/3}$ (see the stress profiles in figure \ref{fig:PointProfiles}).  A similar apparent divergence may be generic in such problems, but is usually cut-off by the finite radius of the indenter (which prevents this divergence reaching the origin). We therefore move on to consider how the effects of a finite indenter size ameliorates this singularity, and the practical relevance, if any, of \eqref{eqn:ForceLawSmallD}.

\section{The role of finite indenter size\label{sec:finite}}

With a finite indenter, $\rin>0$, it is possible to make some analytical progress using the same techniques as outlined in Appendix B for a point indenter. However, in this case there is no analogue of the parametric representation in \eqref{eqn:tdeltaPar}--\eqref{eqn:cFPar}. Instead, we use numerical solutions of the dimensionless problem (see Appendix A), using the MATLAB routine \texttt{bvp4c}. We also consider the limits of small and large indentations asymptotically.

\subsection{Small indentation depths\label{sec:smallindentfiniteRin}}

%
The result for small indentations and a finite indenter radius was given in \eqref{eqn:FLawRin}. Here we merely need to quantify what is meant by `small' in this case:  the force in \eqref{eqn:FLawRin} becomes small compared to that for a point indenter with the same indentation depth $\tdelta$, given by \eqref{eqn:ForceLawSmallD}, when
\beq
\tdelta\sim\tdelta_\ast=4\rin\log(1/\rin).
\label{eqn:FiniteIndentSwitch}
\eeq We therefore expect that for $\tdelta\lesssim \tdelta_\ast$ the force--displacement response is given by \eqref{eqn:FLawRin}. For $\tdelta_\ast\lesssim\tdelta\lesssim1$, however, the details of the indenter are lost and we expect to return to the appropriate result for the point indentation case, namely \eqref{eqn:ForceLawSmallD}.

\subsection{Large indentation depths}

The singular dependence on $\rin$ just observed is coupled to the pre-tension. However, for large indentation depths,  the pre-tension does not play a significant role (as in the point indentation case) and so the behaviour should be well-described by the corresponding solution for a point-indenter.  As a result, we expect to again recover the cubic force-law $\cF\sim\tdelta^3$ with a prefactor  that approaches $\alpha(\nu)$, given in \eqref{eqn:ForceLawLargeD}, in the limit $\rin\to0$.  With $\nu=1/3$, the classic Schwerin solution \cite{Schwerin1929} may readily be adapted to determine $\alpha(\nu;\rin)$ explicitly. In our notation
\beq
\frac{\cF}{\tdelta^3}=\alpha(1/3;\rin)=\frac{\pi}{3}\left(1-\rin^{2/3}\right)^{-3},
\label{eqn:Nu0p33LargeDelta}
\eeq valid for all $\rin<1$. An approximate result corresponding to \eqref{eqn:Nu0p33LargeDelta} for $\nu\neq1/3$, and valid when $\rin\ll1$, is given in Appendix B. 

\begin{figure}
\centering
\includegraphics[width=0.6\columnwidth]{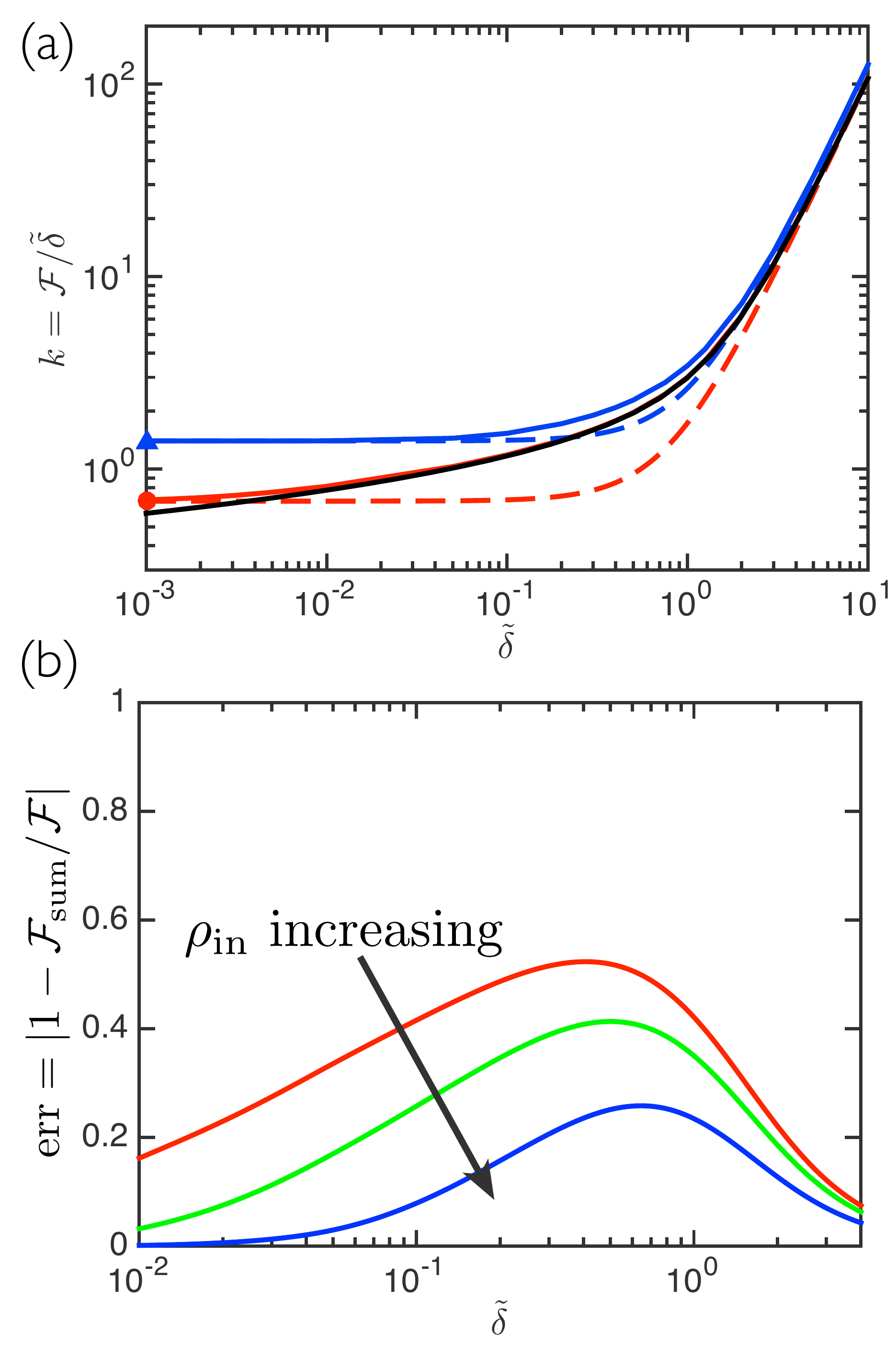}
\caption{The error inherent in using the approximate expression $\Fsum(\tdelta;\rin)$ depends on the size of the indenter and the regime of indentation. (a) The numerically determined stiffness, $k=\cF/\tdelta$, is plotted as a function of $\tdelta$ for a point indenter (black solid curve), $\rin=10^{-4}$ (red solid curve) and $\rin=10^{-2}$ (blue solid curve). This is to be compared to the  approximation, $\Fsum(\tdelta;\rin)$ (defined in \eqref{eqn:simplesum}), which is  shown by the dashed curves of the same colour. For $\tdelta\ll1$, both the numerics and the approximate expression $\Fsum(\tdelta;\rin)$ reproduce the expected constant stiffness mode shown by a circle ($\rin=10^{-4}$) and triangle ($\rin=10^{-2}$). However, at intermediate indentation depths $\tdelta=O(1)$, the error between the approximation and computations grows. (b) The relative error, $\error(\tdelta)$, in the force law $\cF(\tdelta)$ incurred by using $\Fsum(\tdelta;\rin)$, see eqns \eqref{eqn:simplesum} and \eqref{eq:errorSum}. Results are shown for $\rin=10^{-4}$ (red), $\rin=10^{-3}$ (green) and $\rin=10^{-2}$ (blue).
Here $\nu=1/3$ in all computations.
}
\label{fig:ForceLawFinite}
\end{figure}

In the main text we focus on $\nu=1/3$; a comparison between the large indentation result \eqref{eqn:Nu0p33LargeDelta} and numerical simulations is shown in figure \ref{fig:ForceLawFinite}a. This  shows that the force--displacement law is well captured by the asymptotic results \eqref{eqn:FLawRin} and \eqref{eqn:Nu0p33LargeDelta} in the limits of small and large indentation depth, respectively.  Of particular interest is the observation that for small indenters,  $\rin\lesssim 0.01$, the behaviour at intermediate displacements ($4\rin\log(1/\rin)\lesssim \tdelta\lesssim1$) is more accurately described by the point force result \eqref{eqn:ForceLawSmallD},  than the finite indenter result \eqref{eqn:FLawRin}. This observation suggests that experimental results may not actually be in the linear regime \eqref{eqn:FLawRin} for as long as is usually believed. In \S\ref{sec:crit} we will discuss further the implications of this observation.

\subsection{Errors at intermediate displacements\label{sec:errors}}

The results we have discussed thus far hold only for large \emph{or} small displacements. In many practical applications, the range of indentation depths covers an intermediate region. Since no asymptotic results are known that are able to transition smoothly from small to large indentation depths, it is common to form the sum of the small- and large-indentation asymptotic results, giving
\beq
\cF\approx\Fsum=\frac{2\pi}{\log(1/\rin)}\tdelta+\alpha(\nu;\rin)\tdelta^3.
\label{eqn:simplesum}
\eeq In fig.~\ref{fig:ForceLawFinite}(b) we show the relative error introduced by using this simple expression rather than  the true, numerically determined, force law $\cF(\tdelta)$. In particular, we define the relative error
\beq
\error(\tdelta) = |1-\Fsum(\tdelta)/\cF(\tdelta)|.
\label{eq:errorSum}
\eeq We see that the error can in fact be very large for $\tdelta=O(1)$ and, perhaps surprisingly, that this error grows larger as the indenter shrinks. This is due to the fact that as $\rin\to0$, the logarithmic correction to the force, eqn \eqref{eqn:ForceLawSmallD}, becomes important at ever smaller indentation depths.

\section{Indenting a pressurized membrane\label{sec:pres}}

\begin{table*}[h]
\small
  \caption{A summary of the main results for a pre-tensed, pressurized membrane. Note that the results for the force--displacement relations are written in a way that is valid for both small- and large-pressurizations.
}
  \label{tbl:pres2}
  \begin{tabular*}{\textwidth}{@{\extracolsep{\fill}}lllll}
    \hline
   \multirow{2}{*}{Pressurization} & Balloon & Effective & Small & Large \\
   & height & tension & displacements & displacements\\
    \hline
   $\tp\ll1$ & $\ho\sim p\Rf^2/\Tpre$ & $\TpreEff\approx \Tpre$ &&\\
   && & \multirow{2}{*}{$F\approx 2\pi\TpreEff\delta/\log(1/\rin) $} & \multirow{2}{*}{$F/\delta^3 \approx\alpha(\nu;\rin)Y/\Rf^2$} \\
    $\tp\gg1$ & $\ho\sim \left(p\Rf^4/Y\right)^{1/3}$ & $\TpreEff\sim(p\Rf)^{2/3}Y^{1/3}$  &  & \\
    \hline
  \end{tabular*}
\end{table*}

\begin{figure}
\centering
\includegraphics[width=0.3\columnwidth]{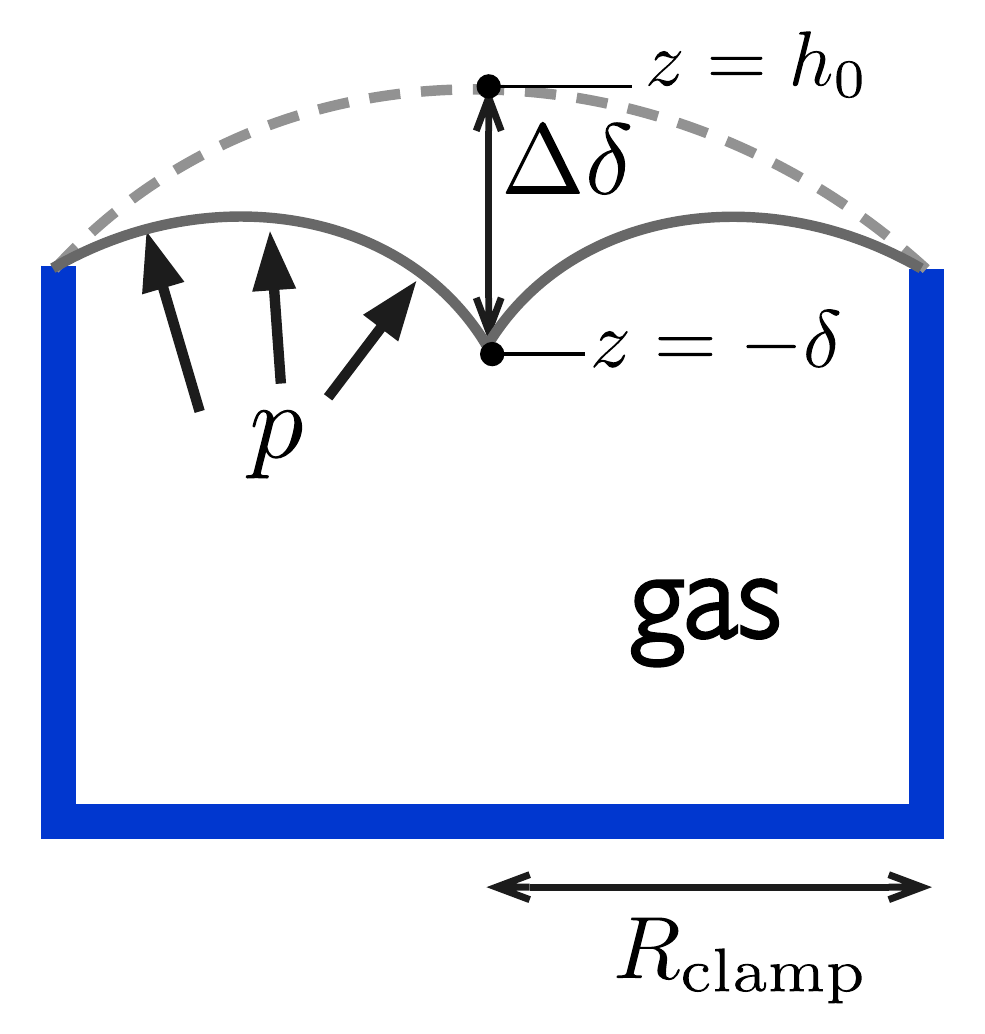}
\caption{Schematic showing the setup for an indented `nano-balloon' (a sheet that is clamped and subject to a pressure difference). In the absence of indentation, the balloon takes a form close to a spherical cap (shown by the grey dashed curve); this cap has height $\ho$, and radius of curvature $\Rcurv\approx \Rf^2/2\ho$. In indentation, the height of the central point is imposed to be a depth $\delta$ below the clamped edges; the indentation depth measured relative to the inflated height is then $\ddelta=\delta+\ho$.
}
\label{fig:Nanoballoon}
\end{figure}

The previous sections  have investigated the effect of the pre-tension and a finite indenter size. However, in many applications the membrane that is being indented is also subjected to a constant pressure difference, $p$, forming a `nano-balloon' (see fig.~\ref{fig:Nanoballoon}). This is particularly relevant for indentation experiments in graphene \cite{LopezPolin2015,LopezPolin2015b,Bunch2008}. We therefore consider next  the effect of a constant applied pressure, $p$, on indentation. 

\subsection{Pressurizing a clamped, pre-tensed membrane}

\begin{figure}
\centering
\includegraphics[width=0.6\columnwidth]{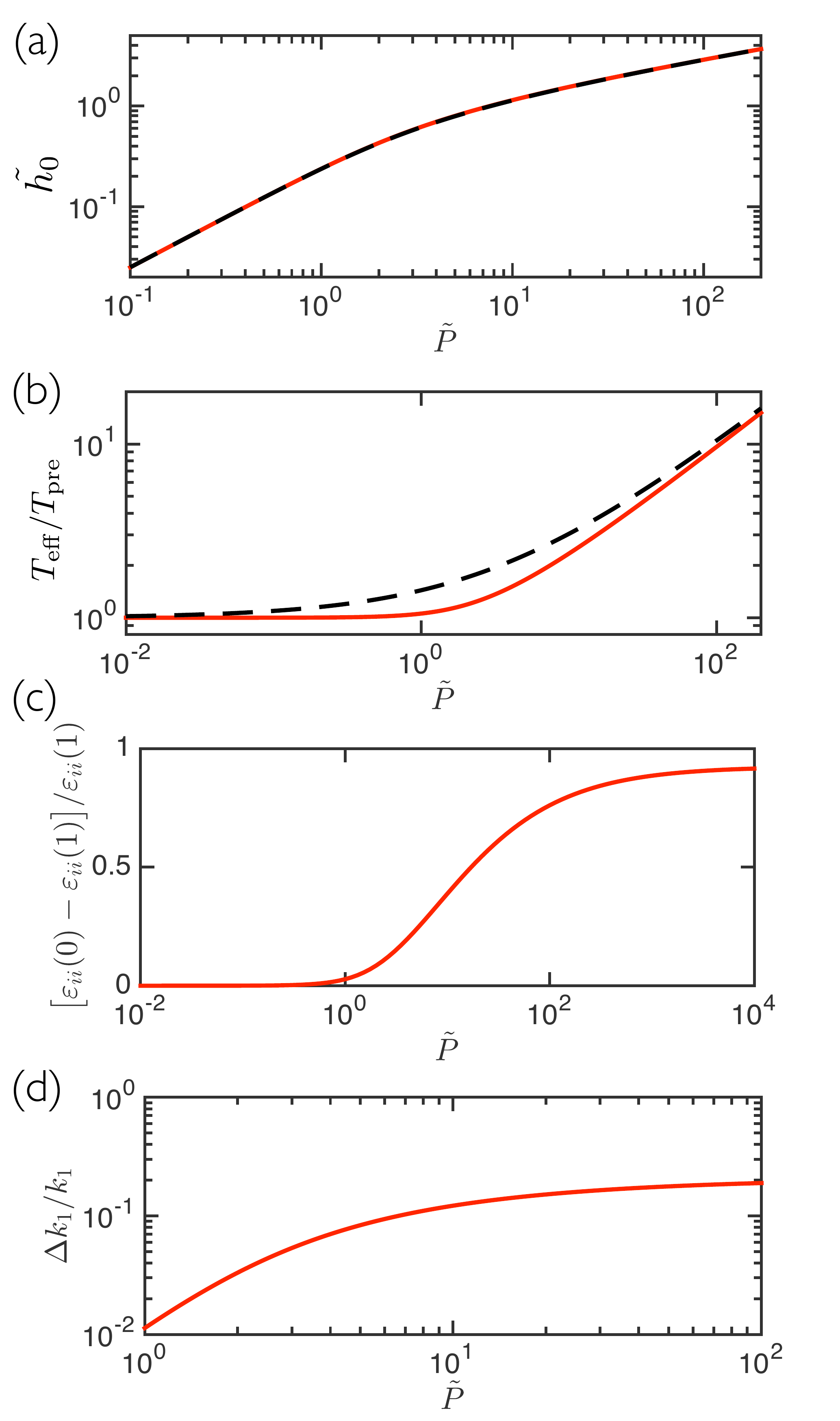}
\caption{The properties of a clamped, unindented and pre-tensed balloon. (a)  The height of the balloon, $\tilde{\ho}=Z(0)$, as a function of the dimensionless pressure $\tp$ (solid curve). The dashed curve shows the approximate analytic result \eqref{eqn:BubbleHeightAsy}. (b) The effective tension at the centre, $\rho=0$, is defined by $\TpreEff=[\srr(0)+\sqq(0)]/2$, and is computed as a function of the dimensionless pressure $\tp$ (solid curve). The dashed curve shows the expression \eqref{eqn:TpreAsy}, which recovers the asymptotic limits $\tp\ll1$ and $\tp\gg1$ correctly but shows significant deviations in-between. (c) The  relative change in the areal strain $\epsilon_{ii}=\epsilon_{rr}+\epsilon_{\theta\theta}$ between the centre of the bubble and the clamped edge.  (d) The relative change in the small indentation stiffness, $k_1=F/\delta$, associated with a $10\%$ change in the inflated bubble height $\ho$. Here all results are obtained with Poisson ratio $\nu=1/3$. }
\label{fig:PressurizedBubbleNoPoke}
\end{figure}

Before discussing the indentation of a nano-balloon, we first consider the shape of the balloon itself. A natural approximation is that the balloon surface will adopt a spherical cap shape, with radius of curvature $\Rcurv$. The problem is then to determine the deformed shape, namely the radius of curvature $\Rcurv$ and the height $\ho$ of the deformed membrane (see fig.~\ref{fig:Nanoballoon}). Assuming the stress within the sheet in nearly uniform, Laplace's law suggests that $\Rcurv\sim T/p$ with $T$ the typical stress within the membrane; $T$ in turn  is related to the pre-tension $\Tpre$ and the strain induced by deformation, $\epsilon\sim (\ho/\Rf)^2$ through
\beqst
T\sim \Tpre+Y(\ho/\Rf)^2.
\eeqst  The final piece of the puzzle is the geometrical relationship $\ho\sim \Rf^2/\Rcurv$ and so we have, in scaling terms, that
\beq
T\sim p\Rf^2/\ho\sim \Tpre+Y(\ho/\Rf)^2.
\label{eqn:scaleTpreEff}
\eeq  Which of the two terms on the RHS of \eqref{eqn:scaleTpreEff} dominates depends on the strength of the pressurization, relative to the pre-tension. We find that the relevant dimensionless parameter measuring this balance is
\beqst
\tp=\frac{p\Rf Y^{1/2}}{\Tpre^{3/2}},
\eeqst which has also been referred to as the `confinement' in related problems \cite{King2012}. In the limit of low pressure, $\tp\ll1$, (or high pre-tension) 
\beq
\ho\sim \frac{p\Rf^2}{\Tpre},\quad T\sim \Tpre.
\label{eqn:lowPres}
\eeq For large pressure, $\tp\gg1$ (or small pre-tension) we have
\beq
\ho\sim \left(\frac{p\Rf^4}{Y}\right)^{1/3},\quad T\sim Y^{1/3}(p\Rf)^{2/3}\sim\Tpre\tp^{2/3}.
\label{eqn:highPres}
\eeq Note that the expression for $\ho$ in the limit $\tp\gg1$ is known from previous works on `bulge tests' \cite{Jensen1991,Vlassak1992}. For $\nu=1/3$, these results may be approximately combined\cite{Mitchell2003} to give the dimensionless pressure
\beq
\tp=4\tilde{\ho}+(\tilde{\ho}/0.645)^3,
\label{eqn:BubbleHeightAsy}
\eeq where $\tilde{\ho}=\ho(Y/\Tpre)^{1/2}/\Rf$ is the dimensionless balloon height. This gives a good account of the numerically-determined behaviour (see fig.~\ref{fig:PressurizedBubbleNoPoke}a).

To address the indentation of a nano-balloon, we need some understanding of the tension close to the point of indentation ---  at the centre of the bulge --- which is what we expect indentation to probe. We therefore define $\TpreEff=[\srr(0)+\sqq(0)]/2$, and plot this as a function of $\tp$ in fig.~\ref{fig:PressurizedBubbleNoPoke}b, noting in the process that the simple composite expansion for the effective tension, obtained by naively combining the asymptotic expressions in \eqref{eqn:lowPres} and \eqref{eqn:highPres}
\beq
\frac{\TpreEff}{\Tpre}=1+0.44\tp^{2/3},
\label{eqn:TpreAsy}
\eeq produces a noticeable error for $\tp=O(1)$. We note also that for $\tp\gg1$ the state of stress is neither isotropic nor uniform. This is illustrated in Fig.~\ref{fig:PressurizedBubbleNoPoke}c, which shows the relative change in the areal strain (the relative local change in area) between the centre and edge of the bubble, as a function of the pressure.

\subsection{Indentation}

To model indentation of a clamped, pressurized and pre-tensed membrane, we incorporate the  pressure $p$ in the normal force balance equation \eqref{eqn:FvK1}; integrating once we find
\beq
\psi\frac{\upd \zeta}{\upd r}=\frac{F}{2\pi}-\frac{p}{2}r^2.
\label{eqn:FvK1P}
\eeq The dimensionless version of this equation (see Appendix A) is solved numerically, together with the dimensionless versions of the compatibility of strains, eqn \eqref{eqn:compat}, and the boundary conditions \eqref{eqn:indBC} and \eqref{eqn:stressBC}. Note that the only change required to account for the pressurization is in the normal force balance, \eqref{eqn:FvK1P}. However, in using \eqref{eqn:indBC} as previously, we emphasize that here $\tdelta$ measures the vertical position of the indenter (see fig.~\ref{fig:Nanoballoon}) and hence \emph{not} the indentation depth relative to the height of the pressurized membrane. This relative indentation depth is denoted $\ddelta$, and is defined to be
\beqst
\ddelta=\delta+\ho.
\eeqst The schematic in fig.~\ref{fig:Nanoballoon} shows these different heights. We continue to use $\tilde{~}$ to signify dimensionless vertical distances, as in \eqref{eqn:DispND}.

\begin{figure}
\centering
\includegraphics[width=0.5\columnwidth]{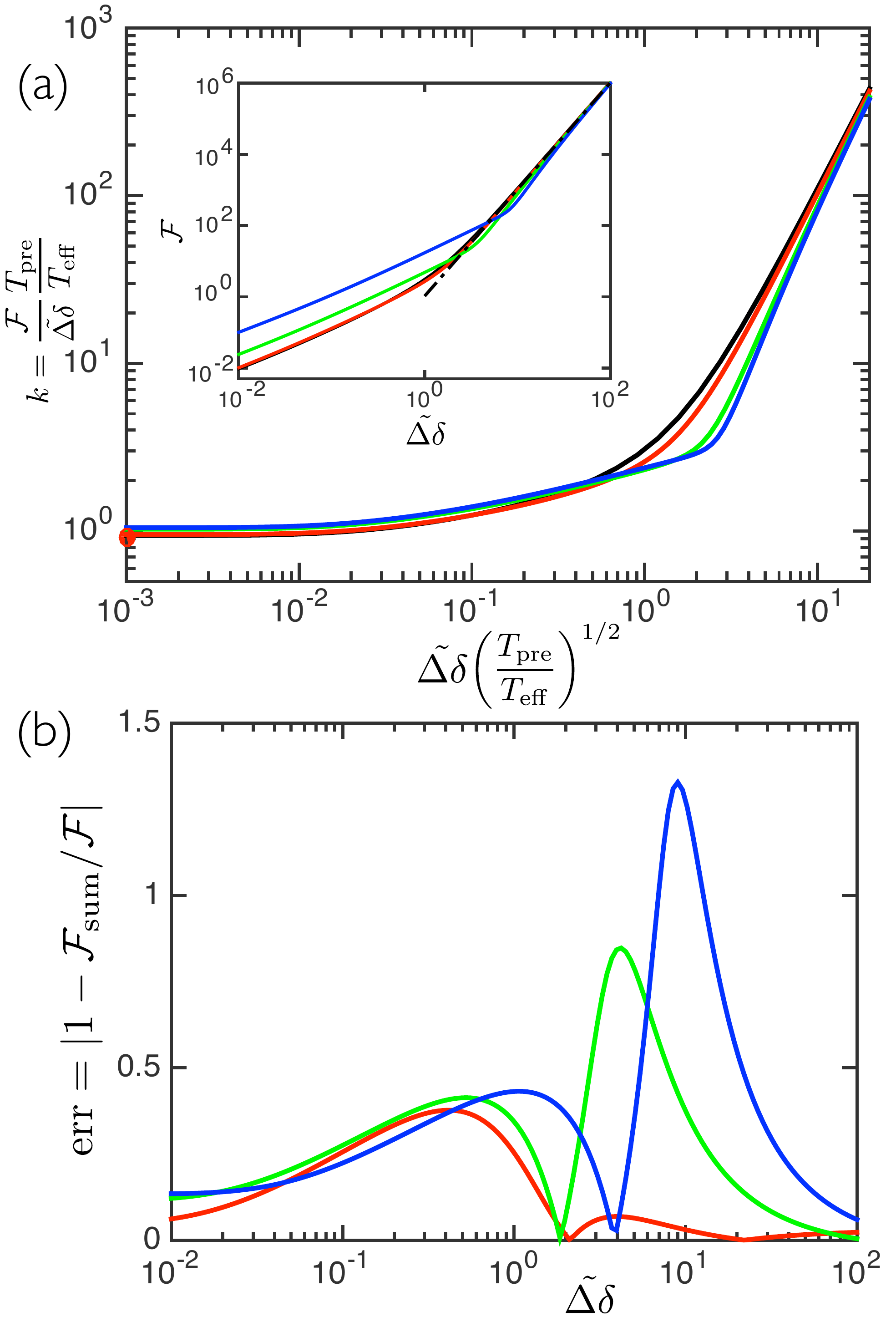}
\caption{The indentation of a pressurized, clamped membrane for various values of the applied pressure $\tp$. (a) The stiffness observed at different pressures can \emph{almost} be rescaled onto a universal curve by using the effective tension, $\TpreEff(\tp)$. This rescaling works well at small and large indentation depths, but fails for at intermediate indentation depths. The inset shows the dimensionless force--indentation curves for different pressures, without any rescaling. (b) The error incurred by using the composite expression \eqref{eqn:PresSum} with $\TpreEff(\tp)$ is enormous at moderate $\tdelta$, and increases with $\tp$. One must therefore ensure that experiments lie in one regime or another, before trying to fit. Here different coloured curves correspond to different extents of pressurization: $\tp=0$ (black) $\tp=1$ (red), $\tp=10$ (green) and $\tp=100$ (blue). (Results with larger $\tp$ are essentially indistinguishable from those with $\tp=100$.) Here, $\nu=1/3$ and $\rin=10^{-3}$ in all computations.
}
\label{fig:PressurizedResults}
\end{figure}

The numerically determined indentation force versus relative displacement, $\dtdelta$, is plotted in the inset of figure \ref{fig:PressurizedResults}(a). Three values of the dimensionless pressurization are used, $\tp=1,10,100$, since these cover the range of behaviours that we observe. At small (relative) indentation depths, $\dtdelta\ll1$, we observe a regime of constant stiffness $\cF/\dtdelta\approx \mathrm{cst}$. As in the unpressurized case, this constant stiffness is caused by the pre-existing tension within the membrane. However, this stiffness now results not from the pre-tension, $\Tpre$, but instead from the pressurization-induced tension $\TpreEff$, which we discussed in scaling terms in \eqref{eqn:scaleTpreEff}. At large indentation depths, $\dtdelta\gg1$, we see that the results tend to the same large indentation asymptote as in the unpressurized case, \emph{i.e.}~$\cF\sim\alpha(\nu)\dtdelta^3$; this again makes intuitive sense since in this limit the indentation-induced stress dominates both the pre-tension and the pressure-induced stress\footnote{Note that this force-indentation result is different to that for a true pressurized elastic shell (with a constant intrinsic radius of curvature), which, for large indentations, recovers a constant stiffness mode  \cite{Vella2012,Vella2015b}; this difference occurs because a shell is able to deform over a horizontal length scale of its choosing, whereas the horizontal length scale of deformation here is fixed by the position of clamping $\Rf$.}. 

To understand the force--indentation curves quantitatively, it  is natural to try and remove the dependence on the pressure by using the effective tension, $\TpreEff(\tp)$. Following  eqns \eqref{eqn:DispND} and \eqref{eqn:ForceND}, we rescale $\dtdelta$ by ${\TpreEff}^{1/2}$ and $\cF$ by ${\TpreEff}^{3/2}$ with $\TpreEff(\tp)$ computed numerically (see fig.~\ref{fig:PressurizedBubbleNoPoke}(b)). This rescaling (main panel of fig.~\ref{fig:PressurizedResults}(a)) shows that the force--indentation response at small and large indentations is precisely as would be expected based on the unpressurized problem considered earlier in this paper; the same asymptotic behaviours are found for  $\tp<0$, but are not shown here. At intermediate indentation depths $\dtdelta=O(1)$, however, we see that the behaviour varies greatly depending on the precise value of $\tp$. The rescaled version of the force law (shown in the main panel of fig.~\ref{fig:PressurizedResults}(a)) highlights this difference: as $\tp$ increases, the transition from $\cF\sim\dtdelta$ to $\cF\sim\dtdelta^3$ becomes sharper, with an almost kink-like transition observed for $\tp=100$. The presence of this kink has important implications for the use of interpolating formulae as well. To study this effect, we define the interpolant
\beq
\Fsum(\dtdelta)=\frac{2\pi\TpreEff(\tp)}{\log(1/\rin)}\dtdelta+\alpha(\nu)\dtdelta^3
\label{eqn:PresSum}
\eeq and measure the relative error, as in \S\ref{sec:errors}. The results, shown in fig.~\ref{fig:PressurizedResults}(b), indicate that the error observed at intermediate indentation depths, $\dtdelta=O(1)$, is substantially larger in the pressurized cases than in the unpressurized cases --- the error reaches $150\%$ for $\tp=100$. 

\subsection{How to determine $p$ and $Y$ from small indentations\label{sec:measureNanoBalloon}}

From the error plots in fig.~\ref{fig:PressurizedResults}(b), it is tempting to conclude that any attempt to measure the stretching modulus of thin materials such as  graphene using indentation is doomed to failure: the stretching modulus only plays a key role in the force--indentation response at very large indentation depths, where the forces quickly become so large that it may not be possible to record them using an AFM. (Indeed, we are not aware of any Graphene experiments in which the cubic force--indentation regime has been reached convincingly.) While it may be tempting to use experimental data obtained at intermediate indentation depths, $\dtdelta=O(1)$, we have shown that this is precisely the regime in which using an interpolating formula such as \eqref{eqn:PresSum} will introduce the largest errors. 

Fortunately, if the pressurization is sufficiently large (\emph{i.e.}~$\tp\gg1$), then this conundrum may be resolved without initially knowing the precise pressure: the height of the unindented balloon, $\ho$, together with indentation data at small indentation depths gives enough information for both $Y$ and $p$ to be inferred.  Classic results for the pressurized blister test\cite{Jensen1991,Koenig2011} give $\ho\approx A_{\ho} \left(p\Rf^4/Y\right)^{1/3}$, while the small indentation stiffness $k_1=F/\delta\approx 2\pi\left[\TpreEff\approx A_\tau Y^{1/3}(p\Rf)^{2/3}\right]/\log(1/\rin)$; here the constants  $A_{\ho}\approx0.645$,  $A_\tau\approx0.438$ for $\nu=1/3$. 

We can then use these relationships to show that
\beq
p\approx\frac{\log(1/\rin)}{2\pi A_{\ho}A_{\tau}}\frac{h_0k_1}{\Rf^2}
\label{eqn:Pdet}
\eeq and
\beq
Y\approx \frac{A_{\ho}^2\log(1/\rin)}{2\pi A_{\tau}}\frac{\Rf^2}{\ho^2}k_1.
\label{eqn:Ydet}
\eeq

The results \eqref{eqn:Pdet}--\eqref{eqn:Ydet} are only valid provided that $\dtdelta\ll1$ and $\tp\gg1$. The signature of being in the small indentation regime is that the linear stiffness $k_1=F/\ddelta$ is approximately constant. The signature of being in the large pressure regime, $\tp\gg1$, is more subtle (unless $\Tpre$, $Y$ and $p$ are all known). However, we note that in this regime,  $k_1$ depends sensitively on $p$: if an experiment were in the  $\tp\gg1$ regime, we would expect $k_1$ to vary noticeably when the experiment is repeated with a slightly different pressure. In contrast, if $k_1$ does not change significantly in such an experiment, one would have to conclude that $\tp\ll1$ instead. More concretely, if the pressure were modified so that the bubble height, $\ho$, increases by, say, $10\%$ then the relative change in the stiffness $k_1$ will be $\approx20\%$ if $\tp\gg1$, and negligible otherwise (see figure \ref{fig:PressurizedBubbleNoPoke}d).

Finally, we note that for graphene, typical values of $Y\approx300\mathrm{~N/m}$ and $\Tpre\approx0.5\mathrm{~N/m}$ have been reported\cite{Lee2008}. This means that with a pressure difference $p\approx4\mathrm{~atm}$ (as in ref.~\cite{LopezPolin2015b}) and drumhead radius $\Rf\approx1\mathrm{~\mu m}$ we might expect to have $\tp\gtrsim20$, making the limit $\tp\gg1$ a reasonable approximation.

\section{Critique of previous works\label{sec:crit}} 

Various analytical results have previously been proposed for the indentation of a pre-tensed membrane. These are repeated and used in the literature, with varying degrees of accuracy.  With the results of the previous sections, we are now in a position to consider some of  these works, and to discuss  their strengths and weaknesses.  This section is divided into five subsections in which we highlight some common flaws and subtleties, examining their impact on indentation-based metrology of thin solid films.   

\subsection{Linear response and the role of indenter size\label{sec:crit:small}}

Several papers\cite{Lee2008,Castellanos2012,Castellanos2015,LopezPolin2015} quote a formula for the force--indentation response,
\beq
F=\pi \Tpre\delta+f(\nu) \frac{Y}{\Ro^2}\delta^3 \ , 
\label{eqn:Schwerin}
\eeq
which is often (incorrectly) attributed to Schwerin \cite{Schwerin1929,Begley2004}. We have seen that the linear term in Eq.~(\ref{eqn:Schwerin}) does not correctly describe even the linear response for an indenter with a finite tip radius $\Ri \ll \Rf$, since the correct linear stiffness \eqref{eqn:FLawRin} contains  a logarithmic dependence on $\rin$.  While such logarithmic factors are usually assumed to be small, for an indenter tip that is 1000 times smaller than the clamping radius, the effect can be significant since the relevant factor $-\log(\rin=10^{-3})\approx 7$. 

Previous measurements of the pre-tension in Graphene \cite{Lee2008,Castellanos2012,Castellanos2015} were based on the erroneous form \eqref{eqn:Schwerin}, and used data for which a significant portion seems to be  in the linear regime ({\emph{i.e.}} where $F \propto \delta$, see Fig.~2A of ref.~\cite{Lee2008}). The neglect of the appropriate logarithmic factor therefore calls into question the validity of the resulting estimate of pre-tension. Other workers in the field \cite{Jennings1995,Wan2003,Tanizawa2004} have correctly appreciated the need for a logarithmic correction, dependent on the tip radius, in the linear regime. One example is the work of Bunch \emph{et al.} \cite{Bunch2008} in which the pre-tension of Graphene was specifically addressed using the correct linear response (Eq.~\eqref{eqn:FLawRin}).  

We also pointed out (see Fig.~4), that when using a small indenter tip one should be careful to use the linear force law Eq.~\eqref{eqn:FLawRin} only for sufficiently small indentations $\tdelta < \tdelta_\ast=4 \rin \log(1/\rin)$; for $\tdelta\gtrsim\tdelta_\ast$ the sub-linear, point-indentation formula, Eq.~\eqref{eqn:ForceLawSmallD}, should be used instead. One potentially useful feature of the intermediate displacement force-law \eqref{eqn:ForceLawSmallD} is that it is not sensitive to the indenter size, provided that  $\tdelta> \tdelta_\ast$. As such, we expect that this result may be applicable even in scenarios in which the effective indenter size changes with indentation depth (\emph{e.g.}~a sphere contacting a membrane will have a contact radius that grows with $\tdelta$).

\subsection{An analytical force-displacement formula} 

An intractable problem with the use of Eq.~(\ref{eqn:Schwerin}) is the assumption that the behaviour of $\cF(\tdelta)$ at intermediate indentation depths, $\tdelta=O(1)$, may be approximated as a sum, $\Fsum(\tdelta)$, of the appropriate asymptotic results in the regimes of small and large indentation, $\tdelta\ll1$ and $\tdelta\gg1$. In \S\ref{sec:errors} we investigated the error (as a function of $\tdelta$) inherent in using such an approximation (with the corrected small indentation behaviour).  We found that although the relative error tends to zero as both $\tdelta \to 0$ and $\tdelta \to \infty$, it may become appreciable for intermediate values of $\tdelta$. This is particularly important since experimental data is often gathered in an intermediate range ({\emph{e.g.}} $ 10^{-2} <\tdelta < 10$). Furthermore, we found that the maximum relative error that is introduced by using such an approximation \emph{grows} as the indenter size shrinks, $\rin\to0$: for $\rin=10^{-3}$, the maximal error is $\sim 40\%$ while for $\rin=10^{-4}$ it is $\sim 50\%$ (see fig.~\ref{fig:ForceLawFinite}(b)). Moreover,  the peak in $\error(\tdelta)$ is not only large, but also broad, affecting a large range of indentation depths.

\begin{figure}
\centering
\includegraphics[width=0.6\columnwidth]{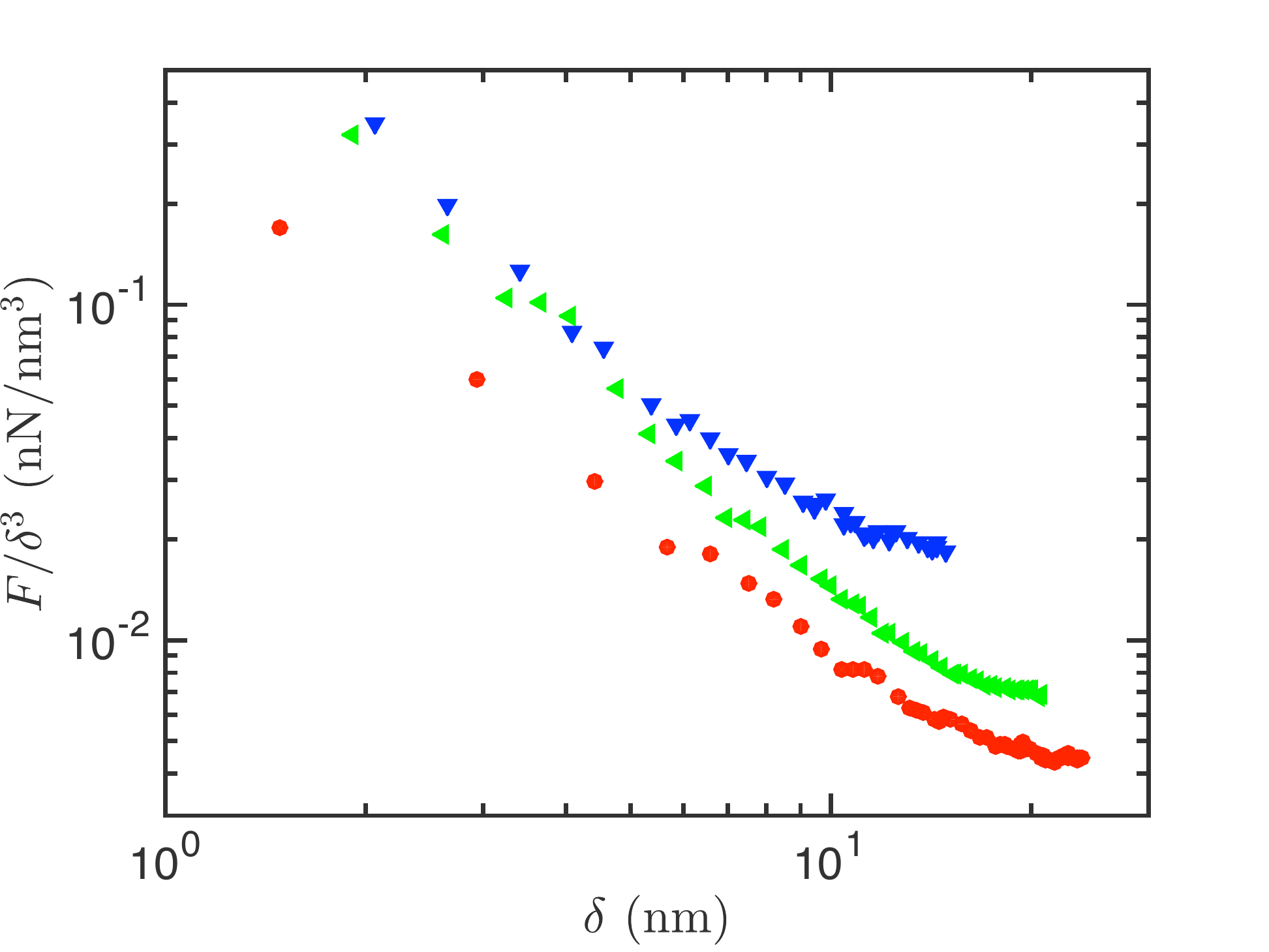}
\caption{Experimental data obtained from the indentation of few-layer mica flakes \cite{Castellanos2015} show that the indentation depth is typically not large enough to reach the regime in which $F/\delta^3=\mathrm{cst}$. Here data are presented for a 2-layer flake (circles), a 3-layer flake (sideways triangles) and a 6-layer flake (downwards triangle).  (The data presented here were captured digitally from fig.~3(c) of ref.~\cite{Castellanos2015}.)}
\label{fig:GomezData}
\end{figure}

Numerous estimates of the stretching modulus of Graphene  \cite{Lee2008,LopezPolin2015} have employed an expression analogous to $\Fsum(\delta)$; as we have shown such approaches are vulnerable to errors at intermediate $\tdelta$. This hurdle may be overcome by using the full numerical solution (solid curves in Fig.~\ref{fig:ForceLawFinite}(a)). Alternatively, if data is available at a sufficiently large range of $\tdelta$, one may simply ignore the data at intermediate $\tdelta$ (signified by a scaling law other than $\cF\propto\tdelta$ or $\cF\propto \tdelta^3$). The data at small indentation depths (signified by $\cF\propto\tdelta$) could then be used to extract the pre-tension (as was done by Bunch \emph{et al.}~\cite{Bunch2008}), while the data at large indentation depths (signified by $\cF\propto\tdelta^3$) could be used to extract the stretching modulus. We emphasize that these two measurements are determined independently of one another, provided that data sits clearly in one of the two separate asymptotic regimes. In several previous studies it appears that experimental data have been used in such a fit despite not lying in the appropriate regime. For example, experiments on few-layer flakes of mica\cite{Castellanos2012,Castellanos2015} were fitted using the expression  \eqref{eqn:Schwerin}. However, reanalysing this data by plotting $F/\delta^3$ shows  that  these experiments do not reach large values of $\delta$: $F/\delta^3$ never reaches the expected plateau (see fig.~\ref{fig:GomezData}).

\subsection{Nano-balloons}

\begin{figure}
\centering
\includegraphics[width=0.6\columnwidth]{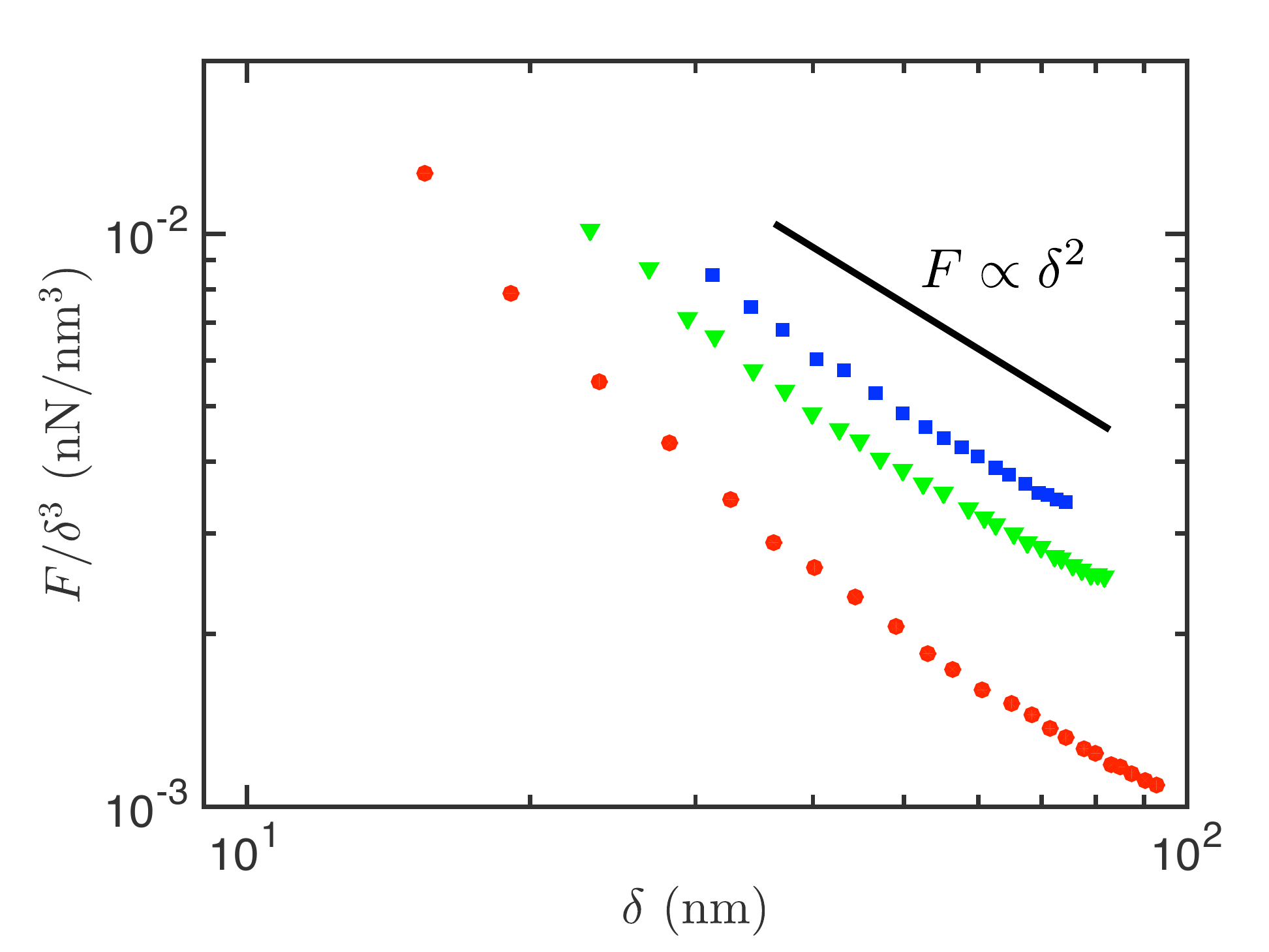}
\caption{Experimental data on the indentation of a graphene nano-balloon from ref.~\cite{LopezPolin2015b}.  $F/\delta^3$ is plotted to show that these data do not reach the large indentation regime $F/\delta^3=\mathrm{cst}$. Here the balloon pressure is varied but not measured directly; instead a globally averaged strain is measured experimentally \cite{LopezPolin2015b}. The reported values of this strain  are $\epsilon=0.09\%$ (circles), $\epsilon=0.23\%$ (triangles) and $\epsilon=0.3\%$ (squares). (The data presented here were captured digitally from fig.~2(b) of ref.~\cite{LopezPolin2015b}.)}
\label{fig:GuineaData}
\end{figure}

Our analysis of indentation in the presence of an internal pressurization has revealed  the perils of using a polynomial fit such as $\Fsum(\tdelta)$: the inaccuracies introduced by using such an expression can be even larger than just discussed, because the `kink' (fig.~\ref{fig:PressurizedResults}a) between linear and cubic behaviour is generally much sharper in this case. (More discussion of this is given in  Appendix C where we  show that a cubic fit of data can lead to large errors in the inferred stretching modulus.) We therefore advise that the first step in any fitting analysis of an indented nano-balloon experiment should be to determine whether the data lies cleanly in one asymptotic regime or another. Perhaps the simplest way of doing this is to use a logarithmic plot of force versus indentation depth, since this will reveal the presence, or lack, of clear power-law behaviour. For example, fig.~\ref{fig:GuineaData}, shows (digitally captured) data from ref.~\cite{LopezPolin2015b}. These suggest power law behaviour that is closest to $F\sim \delta^2$ (and not linear or cubic behaviour). As such, we suggest that these experiments also did not reach large enough indentation depths to reliably extract the stretching modulus $Y$: the experiments are between the small and large indentation regimes, which is precisely where the effect of the switch-over matters. However, we also proposed (see \S\ref{sec:measureNanoBalloon}) that it may be possible to extract the value of $Y$ by focussing instead on the small indentation, large pressure regime, together with measurements of the unindented balloon's height. 

\subsection{Extracting pre-tension from shape}

In a recent paper \cite{Xu2016} it was suggested that an accurate estimate of the pre-tension can be obtained 
by fitting the {\emph{shape}} of the deformed membrane to that predicted by  numerical solutions of the FvK equations. 

\begin{figure}
\centering
\includegraphics[width=0.6\columnwidth]{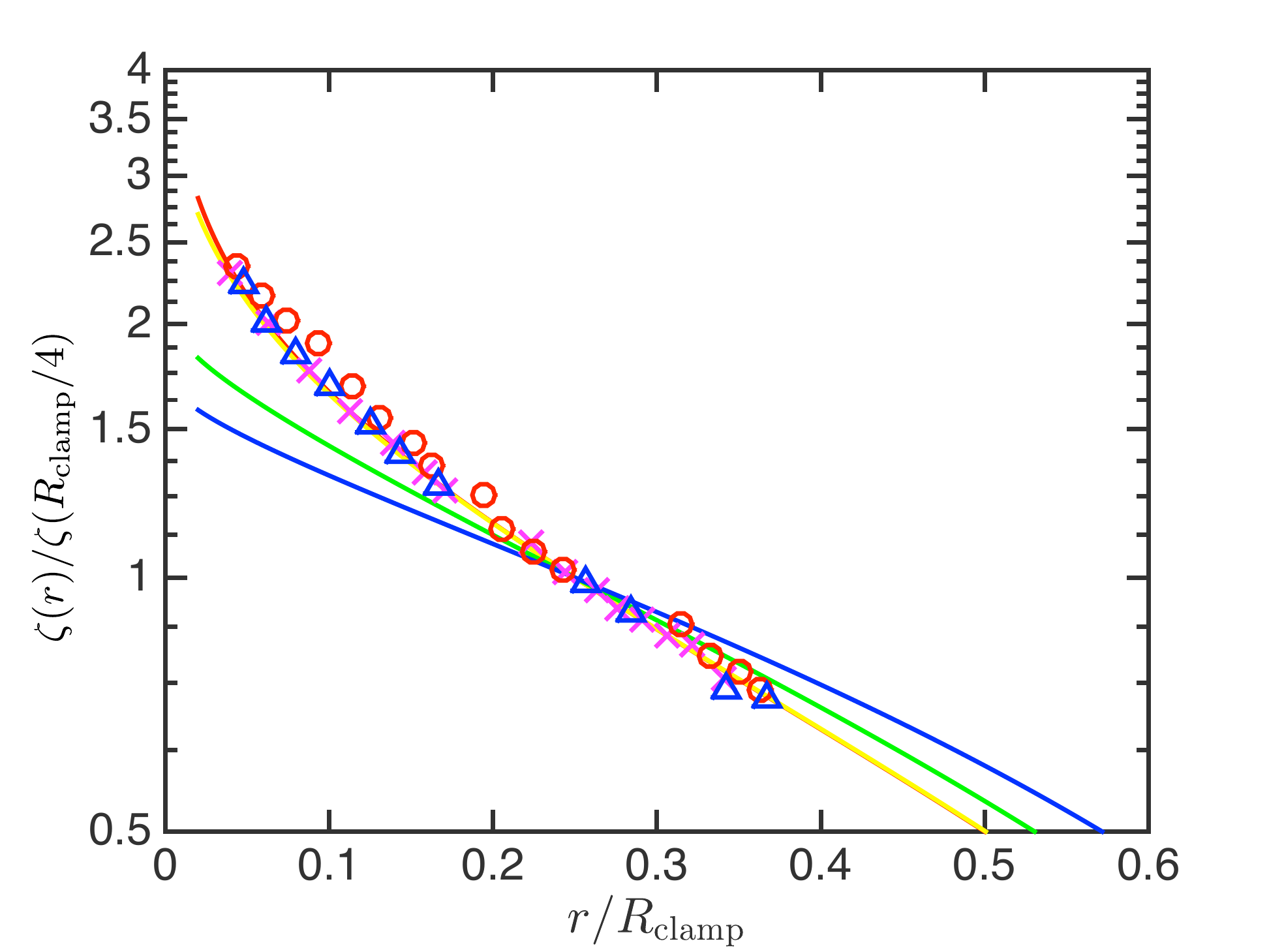}
\caption{Rescaled membrane deflections predicted by our numerical simulations with $\rin=0.02$, $\nu=0.5$ and increasing dimensionless indentation depth: $\tdelta=0.01$ (red curve), $\tdelta=0.1$ (yellow curve), $\tdelta=1$ (green curve) and $\tdelta=10$ (blue curve). A plot with $\tdelta=100$ is indistinguishable from that with $\tdelta=10$ at this scale. The points show experimental data, captured digitally from fig.~3b of ref.~\cite{Xu2016}; we use the same colours to represent data captured at different indentation depth as in figure 3b of ref.~\cite{Xu2016}.)}
\label{fig:MembraneShapes}
\end{figure}

The limitations of this idea can  readily be realized by considering the membrane shapes that are predicted by numerical solutions of the problem. In an experiment, it is not known \emph{a priori} whether a particular indentation depth corresponds to $\tdelta\ll1$ or $\tdelta\gg1$; similarly, the precise value of the indenter radius $\rin$ may not be known (for example, if a small spherical indenter is used\cite{Xu2016}). Figure \ref{fig:MembraneShapes} therefore shows the membrane shapes for different values of $\tdelta$ normalized by the vertical deflection at $\rho=1/4$, \emph{i.e.}~$r=\Rf/4$.  Our numerical solutions show that, when rescaled in this way, the shapes ``collapse" onto two distinct universal shapes corresponding to the linear ($\tdelta \ll 1$) and cubic ($\tdelta \gg 1$) responses. A dependence on $\tdelta$ (and thereby on the pre-tension $\Tpre$) becomes noticeable only within the intermediate parameter range, $\tdelta \sim O(1)$.

We illustrate the importance of these universal shapes by replotting previous experimental results, reproduced from ref.~\cite{Xu2016} and rescaled in precisely the same way (\emph{i.e.}~rescaling $r$ by $\Rf$ and $\zeta(r)$ by $\zeta(\Rf/4)$). These data are shown, together with our numerically determined shapes,  in fig.~\ref{fig:MembraneShapes}. We see that these experimental results are essentially indistinguishable from the numerically determined shapes with $\tdelta\ll1$. We therefore argue that all that can be concluded from such a plot is that in these experiments $\tdelta\lesssim1$ --- any attempt to infer a precise value of $\tdelta$, and hence the pre-tension $\Tpre$, must be subject to so much noise as to be essentially meaningless. 

\subsection{The negligible effect of bending\label{sec:bend}}

Our approach in this study was based on ``membrane theory" in which the bending force is neglected. This approach is a very useful simplification to the analysis, since the $1^{st}$ FvK equation \eqref{eqn:FvK1} is a second-order (rather than fourth-order) differential equation, allowing  analytical progress to be made.

However, other workers \cite{Xu2016,Wan2003} have opted  to retain the bending force in their numerical analysis. In Appendix D we discuss the role of these effects more fully. Here,  we note that the role of bending is expected to be confined to small regions, boundary layers, near the indenter and the outer edge of the film (the regions in which the curvature is largest). In the main portion of the film, the force balance expressed by the simplified membrane theory, \eqref{eqn:FvK1}, must still hold and so we do not expect the force--displacement relationships discussed here to be significantly modified. 

Furthermore, we note that including the bending force requires one to specify additional boundary conditions (such as the slope or torque at the point where the sheet detaches from the indenter). Such boundary conditions are not well-controlled and often need to be  introduced as an additional fitting parameter (see for example ref.~\cite{Xu2016}), adding further uncertainty to the analysis. As such we suggest that the effect of bending can be safely neglected (for sufficiently thin sheets) and, in fact, that this neglect will in general strengthen the robustness of any fitting results that are obtained.

\section{Conclusion\label{sec:conclude}}

Our detailed analysis of the F\"{o}ppl-von-K\'{a}rm\'{a}n equations applied to indentation problems suggests a number of important take-home messages that we summarize here:

\begin{itemize}

\item Using polynomial expressions to fit the measured force, \emph{e.g.}~Eqs.~\eqref{eqn:simplesum} or \eqref{eqn:PresSum}, may lead to erroneous results. This is particularly relevant if data is gathered at an intermediate range of indentation depth, which does not reach the expected cubic regime ($F \sim \delta^3$). Furthermore, the  error inherent in such a fitting increases as the indenter's radius decreases.

\item The errors induced by using a polynomial fit become even more significant when a large pressure difference exists between the two sides of the indented sheet.  This observation undermines attempts to use a cubic fit to extract the stretching modulus from force-deflection data at intermediate indentation depths (see Appendix C for further discussion of this point). 

\item We proposed an alternative method to extract the stretching modulus from data obtained at small indentation depth (\S\ref{sec:measureNanoBalloon}). The caveat of this method is the necessity to achieve sufficiently large pressurization to overcome the pre-tension. We presented a self-consistent test that enables one to verify whether the exerted pressure in the experiment is sufficiently large to allow such an analysis. 

\item We showed that any attempt to obtain metrological data from fitting the shape of the indented sheet (rather than force Vs. deflection) is liable to be extremely inaccurate.  

\end{itemize}

Finally, let us mention that our study was predicated on the assumption that the sheet is strongly clamped at the rim ($r=\Rf$): the radial displacement at the edge of the circular hole is fixed, and does not vary upon indentation. This assumption is implicitly used by most workers in the field, but we are not aware of robust, independent tests of its validity. One case where this assumption may be violated is in pressurized sheets, where an annulus near the rim may be detached from the substrate (if the pressure is ``pushing upward", as in Fig.~\ref{fig:Nanoballoon}) or attached to the wall (if pressure is ``pulling downward"). To account for such a situation, one may have to define an effective clamping radius $\Rf$, which may be slightly smaller than the actual radius of the hole. A more basic subtlety is the assumption that the sheet cannot slide on the substrate. In particular, graphene may be expected to slide easily on sufficiently smooth substrates (due to the weakness of tangential stresses), and so it seems plausible that upon indentation, the sheet will slide inward to reduce the radial stress induced by the indenter. Such a sliding may result in azimuthal (hoop) compression, which can be relaxed by radial wrinkles, thereby affecting substantially the response \cite{Vella2015a}. The consequence of such a scenario will be discussed elsewhere \cite{PacoInPrep}. 
       
\subsection*{Acknowledgments}

We are grateful to Francisco Guinea and Scott Bunch for thoughtful comments on an earlier draft of this manuscript. The research leading to these results has received funding from  the European Research Council under the European Union's Horizon 2020 Programme / ERC Grant Agreement no.~637334 (DV) and NSF-CAREER Grant No. DMR 11-51780 (BD). We also acknowledge support from the W. M. Keck Foundation.

\section*{Appendix A: Dimensionless equations}

In this Appendix we give the complete dimensionless problem (including the applied pressure $\tp$ for completeness). Upon non-dimensionalizing the governing equations \eqref{eqn:FvK1P} and \eqref{eqn:compat} according to \eqref{eqn:NonDim} we find that
\beq
\Psi\frac{\upd Z}{\upd\rho}=\frac{\cF}{2\pi}-\frac{\tp}{2}\rho^2.
\label{eqn:FvK1ndP}
\eeq
and
\beq
\rho\frac{\upd}{\upd \rho}\left[\frac{1}{\rho}\frac{\upd}{\upd \rho}\left(\rho\Psi\right)\right]=-\tfrac{1}{2}\left(\frac{\upd Z}{\upd \rho}\right)^2.
\label{eqn:FvK2nd}
\eeq These are to be solved with the dimensionless version of the boundary conditions \eqref{eqn:indBC} and \eqref{eqn:stressBC}, which are
\beq
\rin\Psi'(\rin)-\nu\Psi(\rin)=(1-\nu)\rin, \quad Z(\rin)=-\tdelta
\label{eqn:bcin}
\eeq and
\beq
\Psi'(1)-\nu\Psi(1)=1-\nu, \quad Z(1)=0.
\label{eqn:bcout}
\eeq These are the equations that are solved numerically in the main text to determine the force--displacement relationship, $\cF(\tdelta)$. 

\subsection*{Small indentations}

In the limit of small indentation depths, $\tdelta\ll1$, and no pressurization, $\tp=0$, we expect that the stress is barely changed from that existing prior to indentation, i.e.~$\Psi\approx\rho$. Substituting this into the vertical force balance equation \eqref{eqn:FvK1ndP} we  obtain
\beqst
\frac{\upd Z}{\upd \rho}=\frac{\cF}{2\pi\rho},
\eeqst which can be integrated subject to the boundary condition \eqref{eqn:bcout} to give
\beqst
Z(\rho)=\frac{\cF}{2\pi} \log\rho.
\eeqst Finally, requiring that $Z(\rin)=-\tdelta$ gives the force law \eqref{eqn:FLawRin}. We also note from \eqref{eqn:FvK2nd} that the correction to $\Psi\approx\rho$ should be expected to enter at $O(\tdelta^2)$.

\section*{Appendix B: Analytical calculation for point indentation}

In this Appendix, we consider the point indentation problem with no pressure, i.e.~$\rin=0$ and $\tp=0$. The dimensionless FvK equations \eqref{eqn:FvK1ndP}--\eqref{eqn:FvK2nd} are to be solved subject to the boundary conditions \eqref{eqn:bcin}, which simplifies to $\Psi(0)=0$, and \eqref{eqn:bcout}.
 

\subsection*{Analytical solution}

We use \eqref{eqn:FvK1ndP} (with $\tp=0$) to eliminate $Z$ from \eqref{eqn:FvK2nd}, giving
\beq
\rho\frac{\upd}{\upd \rho}\left[\frac{1}{\rho}\frac{\upd}{\upd \rho}\left(\rho\Psi\right)\right]=-\tfrac{1}{2}\left(\frac{\cF}{2\pi\Psi}\right)^2.
\label{eqn:FvK2prime}
\eeq At this point it proves useful\cite{Bhatia1968,Chopin2008} to let
\beqst
\eta=\rho^2,\quad \Phi=\rho\Psi
\eeqst so that \eqref{eqn:FvK2prime} becomes
\beq
\frac{\upd^2\Phi}{\upd \eta^2}=-\frac{1}{32\pi^2}\frac{\cF^2}{\Phi^2}, 
\label{eqn:FvK3}
\eeq which is to be solved with boundary conditions
\beq
\Phi(0)=0,\quad 2\Phi'(1)-(1+\nu)\Phi(1)=1-\nu.
\label{eqn:PhiBCs}
\eeq

We can immediately integrate \eqref{eqn:FvK3} once to obtain
\beqst
\frac{\upd\Phi}{\upd \eta}=\frac{\cF}{4\pi}\left(\frac{1+A\Phi}{\Phi}\right)^{1/2}.
\eeqst This can be simplified slightly by letting $\tPhi=A\Phi$ to give
\beqst
\frac{\upd\tPhi}{\upd \eta}=\frac{\cF A^{3/2}}{4\pi}\left(\frac{1+\tPhi}{\tPhi}\right)^{1/2}.
\eeqst

Integrating again, we have that
\begin{align}
\frac{\cF A^{3/2}}{4\pi}\eta&=\int_0^{\tPhi}\left(\frac{f}{1+f}\right)^{1/2}~\upd f\nonumber\\
&=\tPhi^{1/2}(1+\tPhi)^{1/2}-\sinh^{-1}(\tPhi^{1/2})\nonumber.
\end{align} This gives an equation relating the integration constant $A$ and $\cF$ in terms of $\tPhi_1=\tPhi(1)$, which will be useful as a parameter (we will write our analytical solution in terms of the parameter $\tPhi_1$). In particular, we have that
\beq
\cF A^{3/2}=4\pi\left[\tPhi_1^{1/2}(1+\tPhi_1)^{1/2}-\sinh^{-1}(\tPhi_1^{1/2})\right].
\label{eqn:cFparAppend}
\eeq

We can obtain a further equation from the second of the boundary conditions \eqref{eqn:PhiBCs}, which gives an equation for $A(\tPhi_1)$:
\beq
A(1-\nu)=\frac{\cF}{2\pi} A^{3/2}\left(\frac{1+\tPhi_1}{\tPhi_1}\right)^{1/2}-(1+\nu)\tPhi_1,
\label{eqn:AparAppend}
\eeq (since $\cF A^{3/2}$ was already specified as a function of $\tPhi_1$).

Finally, we need to relate the indentation depth $\tdelta$ to $\tPhi_1$. To do this we note that
\begin{equation*}
\tdelta=\int_0^1\frac{\cF}{2\pi\Psi}~\upd\rho=\frac{\cF A}{4\pi}\int_0^{\tPhi_1}\frac{1}{\tPhi}\frac{\upd\tPhi}{\tPhi'}
\end{equation*} which immediately gives
\beq
\tdelta=\frac{2}{A^{1/2}}\sinh^{-1}(\tPhi_1^{1/2}).
\label{eqn:tdeltaparAppend}
\eeq With the set of equations \eqref{eqn:cFparAppend}--\eqref{eqn:tdeltaparAppend} we have a parametric form for the displacement and indentation force in terms of $\tPhi_1$. 

The profile of the membrane, $Z(\rho)$, and the Airy stress function, $\Psi(\rho)$, may also be expressed parametrically as
\beq
Z(\tPhi)=\frac{2}{A(\tPhi_1)^{1/2}}\sinh^{-1}\left[ \tPhi_1^{1/2}\left(1+\tPhi\right)^{1/2}-\tPhi^{1/2}\left(1+\tPhi_1\right)^{1/2}\right]
\label{eqn:parZ}
\eeq and
\beq
\Psi(\tPhi)=A(\tPhi_1)^{-1}\frac{\tPhi}{\rho(\tPhi)},
\label{eqn:parPhi}
\eeq where the radial coordinate, $\rho$, is  given in terms of $\tPhi$ by
\beq
\rho(\tPhi)=\left[\frac{\tPhi^{1/2}(1+\tPhi)^{1/2}-\sinh^{-1}(\tPhi^{1/2})}{\tPhi_1^{1/2}(1+\tPhi_1)^{1/2}-\sinh^{-1}(\tPhi_1^{1/2})}\right]^{1/2}.
\label{eqn:parRho}
\eeq


It is instructive to consider the asymptotic limits of small and large indentations to try and understand the behaviour of the above analytical solution.

\subsubsection*{Small indentations}

In the limit $\tPhi_1\gg1$ we have from \eqref{eqn:AparAppend} that
\beqst
A\sim\tPhi_1
\eeqst (including terms from $\cF A^{3/2}$). We then immediately have that
$\cF\sim 4\pi\tPhi_1/A^{3/2}\sim 4\pi\tPhi_1^{-1/2}$ and
\beqst
\tdelta\sim\frac{2\log(2\tPhi_1^{1/2})}{\tPhi_1^{1/2}}\ll1.
\eeqst Hence the limit $\tPhi_1\gg1$ corresponds to small indentation depths, $\tdelta\ll1$. This result seems counter-intuitive at first but is purely a result of the rescaling used to facilitate the solution: note that $\Psi(1)=\Phi(1)=\tPhi_1/A\sim1$ in the limit $\tPhi_1\gg1$ and so, as expected for small indentation depths, the stress is close to the pre-stress.

To obtain the force--displacement relationship, we eliminate $\tPhi_1$ from the last two expressions to give
\beqst
\tdelta\sim\frac{2\log(2\tPhi_1^{1/2})}{\tPhi_1^{1/2}}\sim \frac{\cF}{2\pi}\log(8\pi/\cF).
\eeqst 

\subsubsection*{Large indentations: $\tPhi_1-\tPhi_1^\ast\ll1$~}\quad To be able to reach large indentation depths, $\tdelta\gg1$, \eqref{eqn:tdeltaparAppend} suggests that we should look for values of $\tPhi_1$ for which $A(\tPhi_1)=0$, i.e.
\beq
(\tPhi_1^\ast)^{1/2}(1+\tPhi_1^\ast)^{1/2}-\sinh^{-1}(\tPhi_1^\ast)^{1/2}=\frac{1+\nu}{2}\frac{(\tPhi_1^\ast)^{3/2}}{(1+\tPhi_1^\ast)^{1/2}}.
\label{eqn:TransEqn}
\eeq  Now, $A(0)=0$ so that $\tPhi_1=0$ is always a possibility. However, as $\tPhi_1\to0$, $\tdelta$ remains finite, and so the root $\tPhi_1=0$ does not correspond to large indentation depths. It is a simple matter to show that for $\nu>1/3$  there is another root of \eqref{eqn:TransEqn}, $\tPhi_1^\ast>0$, while for $\nu<1/3$, this other root is negative, $\tPhi_1^\ast<0$.  To reach the regime $\tdelta\gg1$, we must examine the behaviour close to this other root; we therefore perform  the standard expansions for $\tPhi_1-\tPhi_1^\ast\ll1$. In particular, we have directly from \eqref{eqn:tdeltaparAppend} that to leading order
\beqst
A^{1/2}=\frac{2}{\tdelta}\sinh^{-1}(\tPhi_1^\ast)^{1/2}
\eeqst and hence
\begin{align}
\cF&=4\pi\frac{(\tPhi_1^\ast)^{1/2}(1+\tPhi_1^\ast)^{1/2}-\sinh^{-1}(\tPhi_1^\ast)^{1/2}}{A(\tPhi_1^\ast)^{3/2}}\nonumber\\
&\approx\alpha(\nu)\tdelta^3\label{eqn:ForceLawLargeDReal},
\end{align} where
\beq
\alpha(\nu)=\frac{\pi}{4}(1+\nu)\frac{(\tPhi_1^\ast)^{3/2}}{(1+\tPhi_1^\ast)^{1/2}\left[\sinh^{-1}(\tPhi_1^\ast)^{1/2}\right]^3},
\label{eqn:AlphaNu}
\eeq and we have used \eqref{eqn:TransEqn} to simplify the expression for $\alpha(\nu)$.

Unfortunately, it seems that $\tPhi_1^\ast$ must be determined numerically, and hence that the prefactor $\alpha(\nu)$ in \eqref{eqn:ForceLawLargeDReal} must also be determined numerically. 


\subsection*{Perturbative results for $0<\rin\ll1$}

In the limit of large indentations, $\tdelta\gg1$, the effect of $\rin$ is captured by a regular perturbation theory. We find that
\beq
\frac{\cF}{\tdelta^3}=\alpha(\nu;\rin)\approx\alpha_0(\nu)+\frac{6}{[2\pi(1+\nu)]^{1/3}}\alpha_0^{4/3}\rin^{2/3}
\label{eqn:smallRinLargedelta}
\eeq where $\alpha_0(\nu)$ is the corresponding prefactor in the point-loaded limit $\rin=0$, given in \eqref{eqn:AlphaNu}.

\section*{Appendix C: The perils of polynomial fitting}

A slight modification to the fitting of the analytical expression \eqref{eqn:Schwerin} is to fit experimental data using a cubic polynomial
\beq
\cF=\alpha_0+\alpha_1\tdelta+\alpha_2\tdelta^2+\alpha_3\tdelta^3,
\label{eqn:cubic}
\eeq for free parameters $\alpha_i$; an estimate of the stretching modulus $Y$ may then be found by comparing $\alpha_3$ with the  value of $\cF/\tdelta^3=\alpha(\nu)$ that is expected in the large indentation regime. There are two reasons why this approach seems natural: firstly, this accounts for a shift in the origin caused by inflation (think of the difference between $\dtdelta$ and $\tdelta$ in the nano-balloon problem)\cite{LopezPolin2015b}. Secondly, the  inclusion of constant and quadratic terms in \eqref{eqn:cubic} would seem to give additional freedom to capture the sub-cubic behaviour that is observed in the transition between linear and cubic behaviours (when the indentation depth is not strictly large).

We use our numerical solutions of the fully nonlinear equations to investigate how robust this fitting procedure is: we try to understand when the true cubic behaviour is replicated by the cubic behaviour of the fitted cubic. In other words, we ask when does $\alpha_3$ reproduce the true value of $\alpha(\nu;\rin)$ that is obtained in the asymptotic limit $\tdelta\gg1$? We begin by noting that experimental data, such as that shown in figure \ref{fig:GuineaData}, does not always cover a large range of indentation depths, and, in particular, often does not cover a whole decade in $\delta$. Furthermore, one does not know \emph{a priori} whether an experiment has reached the large indentation regime: without knowing $\Tpre$ and $Y$ one cannot tell whether $\tdelta\gg1$ (as required for the cubic regime to hold) or not. What one can tell from an experiment is whether $k=F/\delta$ varies with indentation depth. In our calculations a greater than $10\%$ variation in $k$ suggests that $\tdelta\gtrsim0.1$; we assume that experimental data corresponding to $\tdelta\lesssim0.1$ would give a behaviour in $F/\delta$ that is close enough to constant to be discarded.

\begin{figure}
\centering
\includegraphics[width=0.6\columnwidth]{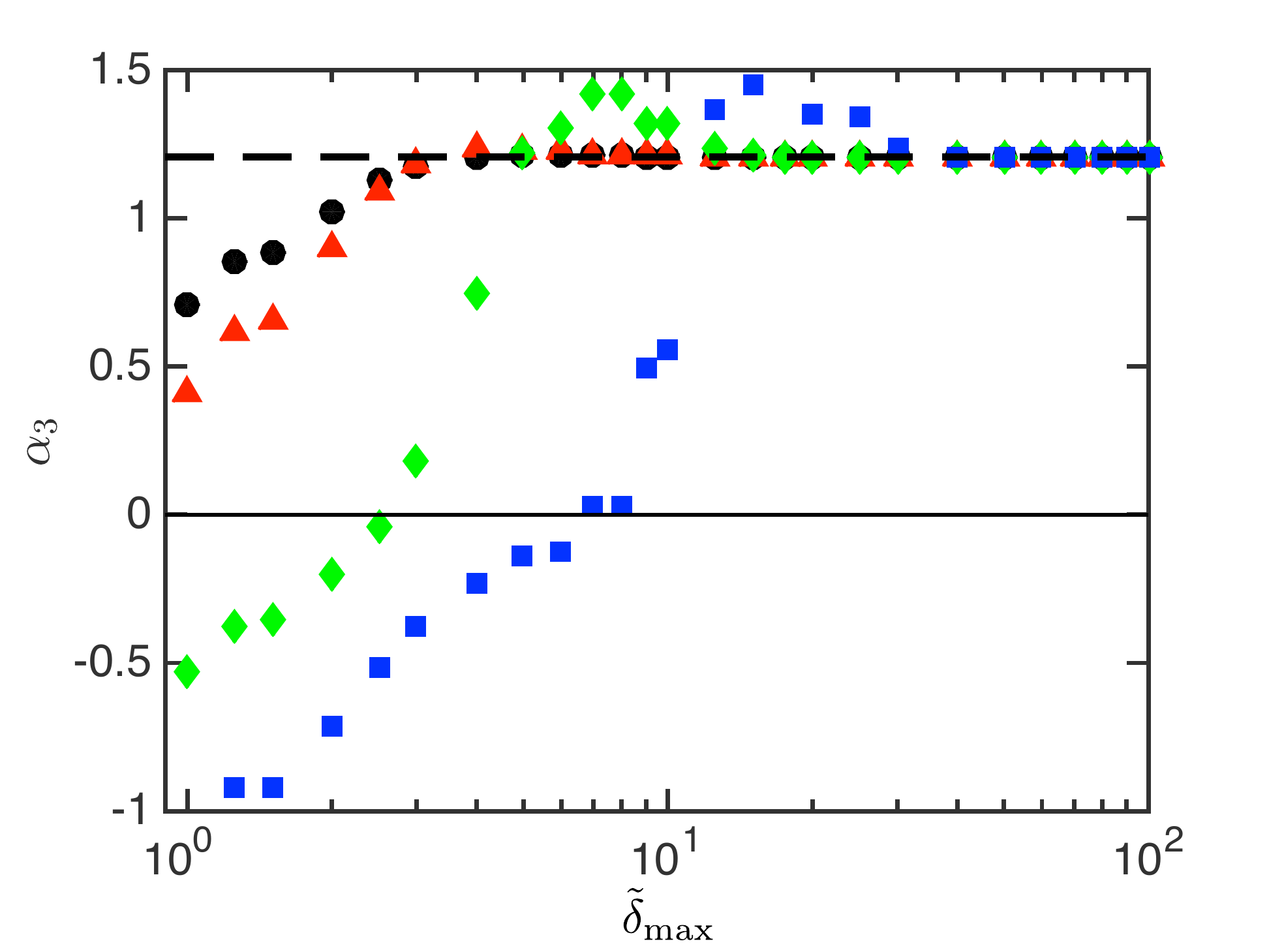}
\caption{How robust is a cubic fit to the range of $\delta$ over which the fit is performed? Here we consider numerically generated data for the indentation of a pressurized balloon over an interval $\tdmax/10\leq\dtdelta\leq\tdmax$ and calculate the coefficient of the cubic term, $\alpha_3$, as this interval changes. Results are shown for $\tp=0$ (circles), $\tp=1$ (triangles), $\tp=10$ (diamonds) and $\tp=100$ (squares).  For sufficiently large $\tdmax$ we recover the expected result, $\alpha_3\to \pi(1-\rin^{2/3})^{-3}/3$, which is shown by the dashed horizontal line. Here $\nu=1/3$ and $\rin=10^{-2}$.}
\label{fig:cubicfit}
\end{figure}

We therefore consider our numerical ``data" restricted to intervals $\tdelta\in[\tdmin,\tdmax]$ with $\dmin=\dmax/10\geq0.1$ taken for definiteness. We then make a cubic fit of this data, cf.~\eqref{eqn:cubic}, and extract the corresponding value of $\alpha_3$ from this fit. Figure \ref{fig:cubicfit} shows the results of this analysis for numerical data generated with $\rin=10^{-2}$ and  internal pressurizations $\tp=0,1,10$ and $100$. We see that $\alpha_3\to\pi(1-\rin^{2/3})^{-3}/3$ as $\tdmax$ grows --- this is as should be expected since this corresponds to the truly nonlinear regime. However, for ranges that cover the intermediate indentation regimes (where the true force law is neither cubic nor linear) we see a large variation in the value of $\alpha_3$. This effect is particularly large for pressurized membranes with large $\tp$; indeed, for large pressurizations and small enough $\tdmin$ it is even possible to find $\alpha_3<0$ through this procedure. Clearly this is an artefact of the fitting procedure, and does not have any physical significance.

This analysis shows that this fitting procedure is actually quite sensitive to the interval on which the fitting is done. In particular, we see that even in the unpressurized limit ($\tp=0$) one could make an error of at least $50\%$ simply by attempting the cubic fit over an inappropriate interval of $\delta$.

\section*{Appendix D: Neglecting bending}

In \S\ref{sec:bend} we discussed our neglect of bending briefly. To justify this, we only need to perform a consistency check: it is enough to evaluate the bending force that would be associated with the membrane profiles that we obtained in sections \ref{sec:point}-\ref{sec:pres}, and compare this force with the indentation force that we calculated for these same profiles. One may easily see that the bending force scales as:
\beq
\Fbend \sim  B \zeta'''' \sim B \delta/\Rf^4   \ , 
\label{eq:bend-estimate1}
\eeq  
where spatial derivatives are estimated based on the clamping radius. The ratio between the ``membrane force" , $\Fmem$, and $\Fbend$, which is often called  the ``bendability" \cite{Davidovitch2011,Hohlfeld2015,Vella2015a}, requires us to identify the dominant membrane force. As we showed in previous sections, this may be induced by pre-tension (for $\tdelta \ll 1$ and $\Tpre \gg (P\Rf)^{2/3} Y^{1/3}$),  pressure (for $\dtdelta \ll1$ and $\Tpre \ll (P\Rf)^{2/3} Y^{1/3}$), or 
the stretching modulus of the sheet (for $\tdelta \gg 1$ or $\tdelta \gg \tilde{\ho}$, respectively, for large and small pretension-to-pressure ratio). Thus the bendability is
\beq
 \frac{\Fmem}{\Fbend} \sim  \max \left\{ \frac{\Tpre \Rf^2}{B} , \frac{P^{2/3} Y^{1/3} \Rf^{8/3}}{B} ,   
\frac{Y \delta^2}{B} \right\} \ . 
\label{eq:bend-estimate2}
\eeq  In most experimental scenarios that we are aware of, the above ratio is $\gtrsim10^4$.

Thus, at the macroscopic scale of the whole film, the effect of bending may be neglected. Nevertheless, the bending force may be relevant at small scales, where the spatial variation is sufficiently rapid, as is typical in ``boundary layers". Here we expect the boundary layers (in the vicinity of the indenter and/or the clamped edge) to have  typical horizontal scale $\lec= \sqrt{B/\sigma}$ (where 
$ \sigma = \max\{  \Tpre,(P\Rf)^{2/3}Y^{1/3}\}$, or $\sim Y(\delta/\Rf)^2$); over the length scale $\lec$ the effect of any exerted torque relaxes. The net contribution of this bending-induced force to the total force is inversely proportional to the bendability and can be safely neglected in most practical situations for very thin sheets. (We note further that in some situations membrane theory may predict compressive stresses; in such scenarios, bending has a strong, non-perturbative effect on the stress field, which eliminates any such compression, see \emph{e.g.}~refs \cite{Davidovitch2011,Davidovitch2012,Schroll2013,Vella2015a}; however, this is not the case for indenting a clamped sheet, where the stresses remain purely tensile everywhere.)

\end{document}